\documentclass[10pt,aps,prd,showpacs,twocolumn,superscriptaddress,nofootinbib]{revtex4-1}
\usepackage{amsfonts,amsmath,amssymb,amsthm}
\usepackage[bookmarks, bookmarksopen, bookmarksnumbered]{hyperref}
\usepackage{graphicx}
\usepackage{tensor}
\usepackage{color}

\begin{document}

\title{Self-gravitating magnetised tori around black holes in general relativity}

\author{Patryk Mach}
\affiliation{Instytut Fizyki im.~Mariana Smoluchowskiego, Uniwersytet Jagiello\'nski, {\L}ojasiewicza 11, 30-348 Krak\'{o}w, Poland}

\author{Sergio Gimeno-Soler}
\affiliation{Departamento de Astronomia y Astrof\'{\i}sica, Universitat de Val\`encia, Dr.~Moliner 50, 46100 - Burjassot, Spain}

\author{Jos\'{e} A.\ Font}
\affiliation{Departamento de Astronomia y Astrof\'{\i}sica, Universitat de Val\`encia, Dr.~Moliner 50, 46100 - Burjassot, Spain}
\affiliation{Observatori Astron\`omic, Universitat de Val\`encia, Jos\'e Beltr\'an 2, 46980, Paterna, Spain}

\author{Andrzej Odrzywo\l{}ek}
\affiliation{Instytut Fizyki im.~Mariana Smoluchowskiego, Uniwersytet Jagiello\'nski, {\L}ojasiewicza 11, 30-348 Krak\'{o}w, Poland}

\author{Micha\l{} Pir\'{o}g}
\affiliation{Institut f\"{u}r Theoretische Physik, Max-von-Laue-Str.~1, D-60438 Frankfurt am Main, Germany}

\begin{abstract}
We investigate stationary, self-gravitating, magnetised disks (or tori) around black holes. The models are obtained by numerically solving the coupled system of the Einstein equations and the equations of ideal general-relativistic magnetohydrodynamics. The mathematical formulation and numerical aspects of our approach are similar to those reported in previous works modeling stationary self-gravitating perfect-fluid tori, but the inclusion of magnetic fields represents a new ingredient. Following previous studies of purely hydrodynamical configurations, we construct our models assuming Keplerian rotation in the disks and both spinning and spinless black holes.  We focus on the case of a toroidal distribution of the magnetic field and build a large set of models corresponding to a wide range of values of the magnetisation parameter, starting with weakly magnetised disks and ending at configurations in which the magnetic pressure dominates over the thermal one. In all our models, the magnetic field affects the equilibrium structure of the torus mainly due to the magnetic pressure. In particular, an increasing contribution of the magnetic field shifts the location of the maximum of the rest-mass density towards inner regions of the disk. The total mass of the system and the angular momentum are affected by the magnetic field in a complex way, that depends on the black hole spin and the location of the inner radius of the disk. The non-linear dynamical stability of the solutions presented in this paper will be reported elsewhere.
\end{abstract}

\maketitle

\section{Introduction}

Compact accretion disks (or tori) around black holes are astrophysical transient systems that can form in a number of situations. Examples include the core-collapse of massive stars~\cite{woosley:2006}, the merger of compact binaries consisting of either two neutron stars or a black hole and a neutron star (see e.g.~\cite{baiotti2017} and references therein), and the gravitational collapse of a supermassive star~\cite{rees1984,SMS}. Observations of the formation and evolution of black hole--torus systems are challenging, either using neutrino, electromagnetic or gravitational-wave approaches. Since the recent breakthrough observation of gravitational waves from a binary neutron star (BNS) merger by Advanced LIGO and Virgo~\cite{GW170817,GW-GRB} one may hope that this cosmic messenger may offer the best possibility of observing black hole--torus systems in the near future.

Numerical relativity is the best approach to study the dynamical formation of black hole--torus systems from \textit{ab-initio} simulations. Long-term simulations of BNS mergers that include the late inspiral of the two neutron stars and account for the relevant physics (i.e.~relativistic gravity, general-relativistic magnetohydrodynamics (GRMHD), and neutrino transport) are in general fairly expensive. Therefore, building equilibrium initial data of black hole--torus systems is highly motivated, as it allows to carry out follow-up studies of the last stages of the merger in a less expensive way and in a more controlled environment, sidestepping the computation of the late inspiral and early merger phase. Equilibrium models must therefore be as faithful as possible to the end-products of the numerical evolutions, increasing their realism as new physical ingredients are incorporated (see~\cite{abramowicz:2013} and references therein). Numerical works have shown that the mass of the tori may be large enough to render necessary to account for the disk self-gravity in order to properly describe its dynamics. This is particularly true for the case of unequal-mass BNS mergers~\cite{baiotti2017,rezzolla10}. Motivated by these results, we present in this paper new families of self-gravitating disks around black holes. 

A few authors have previously investigated this issue~\cite{nishida, ansorg, shibata, stergioulas}. In their seminal work, Nishida and Eriguchi~\cite{nishida} computed self-gravitating toroids around stars and black holes using Komatsu-Eriguchi-Hachisu's (KEH) self-consistent-field method~\cite{komatsu}. Elliptic-type field equations were converted into integral equations using Green's functions. Later on, Ansorg and Petroff~\cite{ansorg} built solutions of black holes surrounded by uniformly rotating rings of constant density using the same approach as~\cite{nishida}, but solving the equations with a highly accurate multi-domain, pseudo-spectral method. A similar strategy was followed by Stergioulas~\cite{stergioulas} to construct general-relativistic models of self-gravitating, constant angular momentum tori around black holes with KEH's self-consistent-field method. An important ingredient of this approach was the use of a compactified radial coordinate, which improved the enforcement of the boundary conditions asymptotically. Among all previous studies,  the most relevant one for our work is that of Shibata~\cite{shibata}, since we follow very closely his procedure. Shibata's work departs from the other three approaches in that it builds self-gravitating tori around rotating black holes adopting the so-called puncture framework to describe the spacetime of a rotating black hole, and hence avoiding potential numerical issues when dealing with the curvature singularity at the origin. The models reported by~\cite{shibata} are purely hydrodynamical (i.e.\ with no magnetic field), and they are characterized by the constant angular momentum. (Non-constant angular momentum tori were considered in~\cite{kiuchi2011} albeit around non-rotating black holes.) On the contrary, the models presented in this paper incorporate a toroidal distribution of the magnetic field, a physically motivated Keplerian rotation law~\cite{kkmmop,kkmmop2} and rotating black holes.

In addition to self-gravity, our disks also incorporate magnetic fields, within the ideal GRMHD approach. To the best of our knowledge, equilibrium sequences of self-gravitating and magnetised disks around black holes in general relativity have not yet been reported in the literature. (Notice, however, that Appendix A of~\cite{shibata} outlines the procedure to build a self-gravitating magnetised disk with a toroidal magnetic field, but no examples are provided.) There exist a number of previous works where equilibrium solutions of magnetised disks around black holes have been built~\cite{komissarov:2006,montero:2007,gf} but, to the best of our knowledge, all of them are restricted to the test-fluid approximation (i.e.~neglecting self-gravity). Komissarov~\cite{komissarov:2006} first presented a general procedure to build magnetised ``Polish doughnuts" (constant angular momentum tori) using a barotropic equation of state and the assumption that  the  specific  enthalpy  of  the  fluid  is  close  to  unity. This restrictive condition on the thermodynamics was relaxed in the work of Montero et al.~\cite{montero:2007}, who also performed dynamical evolutions of those tori. More recently, Gimeno-Soler and Font \cite{gf} built new sequences of equilibrium magnetised tori around Kerr black holes assuming a form of the angular momentum distribution proposed in \cite{qian:2009} that departs from the constant case of \cite{komissarov:2006} and from which the equipotential surfaces can be easily computed. 

The study of the stability of equilibrium solutions of accretion tori under perturbations has received considerable numerical attention (see~\cite{abramowicz:2013} for a review). In particular, constant angular momentum disks have been found to be generically unstable. On the other hand, in most BNS merger simulations the final black hole--torus system does not manifest signs of dynamical instabilities on short dynamical timescales (see~\citep{baiotti2017} and references therein). Specifically, the simulations of \cite{rezzolla10} indicate that the angular velocity $\Omega$ of tori formed from unequal-mass BNS mergers follows Keplerian profiles, $\Omega \propto r^{-3/2}$, where $r$ denotes the distance from the rotation axis, which  explains the scaling of the specific angular momentum as $r^{1/2}$. This provides firm evidence that tori produced self-consistently are dynamically stable. However, despite their non-constant angular momentum profiles make them stable against the development of the so-called runaway instability~\citep{abramowicz,Font2002,montero}, on longer timescales non-axisymmetric instabilities (e.g.~the Papaloizou-Pringle instability (PPI) \citep{PPI}) set in~\citep{korobkin2011,kiuchi2011,mewes16a,mewes16b}. Recently Bugli et al.~\cite{bugli} studied the development of the PPI in tori threaded by weak toroidal magnetic fields and how this instability may be affected by the concurrent development of the magnetorotational instability (MRI). Their simulations, within the test-fluid limit, showed that the magnetic fields provide local viscous stresses through turbulence and global angular momentum transport,  leading to the suppression of large-scale PPI modes. The self-gravitating, magnetised tori we built in the present work may thus be used in the future to investigate the generality of the findings of~\cite{bugli} beyond the test-fluid limit.

The paper is organised as follows: in Sec.~\ref{equations} we discuss the mathematical aspects of our procedure, presenting the Euler-Bernoulli equations, the Einstein equations, and the Keplerian rotation law. The masses and angular momentum of the black hole--torus spacetime are discussed in Sec.~\ref{masses_ang_mom}. Section~\ref{method} briefly describes our numerical method and the results are discussed in Sec.~\ref{results}. Finally, Sec.~\ref{summary} gives a summary of this work. In the Appendix we provide expressions for the Kerr metric in quasi-isotropic coordinates. We use geometric units with $G = c = 1$, where $G$ is Newton's constant and $c$ is the speed of light, and assume the signature of the metric $(-,+,+,+)$. Spacetime dimensions are labeled with Greek indices, $\mu = 0,1,2,3$, while Latin indices are used to denote spatial dimensions, $i = 1,2,3$.

\section{Equations}
\label{equations}

We start by deriving the equations describing stationary, axially symmetric, self-gravitating, magnetised toroids rotating around black holes. The black hole (which can be spinning) is included in the system using the puncture method. The torus is described in terms of ideal GRMHD; we restrict ourselves to toroidal magnetic fields and barotropic equations of state. In specific numerical examples discussed in Sec.\ \ref{results} we assume polytropic fluids and a Keplerian rotation law introduced recently in \cite{kkmmop,kkmmop2}.

As mentioned in the introduction the formulation presented in this paper is based on the approach to modeling stationary perfect-fluid disks around black holes derived originally in \cite{shibata} for the case with no magnetic fields. In particular, in the derivation of the equations we follow closely the steps taken in \cite{shibata}. Because the terms connected with the magnetic field appear irregularly in many of the equations, we repeat the corresponding calculations also in this paper.  The magnetic field enters the description of stationary disks in two places: in the stationary Euler equation and in the ``source terms'' of the Einstein equations. 

Within the framework of ideal GRMHD the energy-momentum tensor has the form
\begin{equation}
\label{emtensor}
T_{\mu \nu} = (\rho h + b^2) u_\mu u_\nu + \left( p + \frac{1}{2} b^2 \right) g_{\mu \nu} - b_\mu b_\nu,
\end{equation}
where $\rho$ is the baryonic density, $h$ is the specific enthalpy, $p$ is the (thermal) pressure, $u^\mu$ denotes components of the four-velocity of the fluid, and $b^\mu$ is the four-vector of the magnetic field. We denote $b^2 = b_\mu b^\mu$. Note that the quantity $p_\mathrm{mag} = \frac{1}{2} b^2$ plays the role of a magnetic pressure. It is assumed that
\begin{equation}
\label{orthogonal}
b_\mu u^\mu = 0.
\end{equation}
In this case the dual of the Faraday tensor relative to an observer with four-velocity $u^\mu$, $^\ast F^{\mu\nu} = b^\mu u^\nu - b^\nu u^\mu$, satisfies $\nabla_\mu {}^\ast F^{\mu \nu} = 0$.

We will work in spherical coordinates $(t, r , \theta, \varphi)$. It is convenient to start with a general, stationary and axially symmetric metric of the form
\begin{equation}
\label{generalmetric}
g = g_{tt} dt^2 + 2 g_{t \varphi} dt d\varphi + g_{rr} dr^2 + g_{\theta \theta} d\theta^2 + g_{\varphi \varphi} d\varphi^2, 
\end{equation}
where the metric potentials $g_{tt}$, $g_{t\varphi}$, $g_{rr}$, $g_{\theta \theta}$, $g_{\varphi \varphi}$ depend on $r$ and $\theta$ only. We consider an axially symmetric, stationary configuration with $u^r = u^\theta = b^r = b^\theta = 0$. It follows from Eq.\ (\ref{orthogonal}) that
\begin{equation}
b_t = - \frac{u^\varphi}{u^t} b_\varphi = - \Omega b_\varphi,
\end{equation} 
where $\Omega = u^\varphi/u^t$. Note that the normalization of the four-velocity $u_\mu u^\mu = -1$ yields
\begin{equation}
\label{ut}
g_{tt} + 2 g_{t \varphi} \Omega + g_{\varphi \varphi} \Omega^2 = - \frac{1}{(u^t)^2}.
\end{equation}
It can be easily shown that
\begin{equation}
b_\varphi^2 = - (u^t)^2 \mathcal L b^2,
\end{equation}
where $\mathcal L = g_{\varphi \varphi} g_{tt} - g_{t \varphi}^2$.

\subsection{Euler-Bernoulli equation}

The way of deriving the Euler-Bernoulli equation (or the first integral of the Euler equations) for the ideal GRMHD energy-momentum tensor is described in Appendix A of \cite{shibata}. The computation of the four-divergence
\begin{equation}
\nabla_\mu T\indices{^\mu_\nu} = \frac{1}{\sqrt{-g}} \partial_\mu \left( \sqrt{-g} T\indices{^\mu_\nu} \right) - \frac{1}{2} (\partial_\nu g_{\alpha \beta}) T^{\alpha \beta} = 0
\end{equation}
yields
\begin{equation}
\label{bernoullideriva}
\partial_\nu \left( p + \frac{1}{2} b^2 \right) - \frac{1}{2}(\partial_\nu g_{\alpha \beta}) \left[ (\rho h + b^2) u^\alpha u^\beta - b^\alpha b^\beta \right] = 0.
\end{equation}
The above equation is trivially satisfied for $\nu = t$ and $\nu = \varphi$. Nontrivial information is contained in Eq.\ (\ref{bernoullideriva}) for $\nu = r, \theta$. Following \cite{shibata}, one can show that
\begin{equation}
\frac{1}{2} u^\alpha u^\beta \partial_\nu g_{\alpha \beta} = \frac{\partial_\nu u^t}{u^t} - u^t u_\varphi \partial_\nu \Omega
\end{equation}
and
\begin{equation}
\frac{1}{2} b^\alpha b^\beta \partial_\nu g_{\alpha \beta} = b^2 \left( \frac{\partial_\nu u^t}{u^t} + \frac{\partial_\nu \mathcal L}{2 \mathcal L}  - u^t u_\varphi \partial_\nu \Omega \right).
\end{equation}
Combining the above expressions one gets
\begin{equation}
\rho h \left( u^t u_\varphi \partial_\nu \Omega - \frac{\partial_\nu u^t}{u^t} \right) + \partial_\nu p + \frac{\partial_\nu (b^2 \mathcal L)}{2 \mathcal L} = 0,
\end{equation}
or, dividing by $\rho h$,
\begin{equation}
u^t u_\varphi \partial_\nu \Omega - \frac{\partial_\nu u^t}{u^t} + \frac{\partial_\nu h}{h} + \frac{\partial_\nu (b^2 \mathcal L)}{2 \rho h \mathcal L} = 0,
\end{equation}
where we have used the fact that $dh = dp/\rho$. Therefore, it is possible to search for a solution in the form
\begin{equation}
\int u^t u_\varphi d\Omega + \ln \left( \frac{h}{u^t} \right) + \int \frac{d(b^2 \mathcal L)}{2 \rho h \mathcal L} = C,
\end{equation}
which is Eq.\ (A11) of \cite{shibata} (note a misprint in the last term of the equation given in \cite{shibata}). The above equation is also equivalent to Eq.\ (11) of \cite{gf}. We define the angular momentum per unit inertial mass $\rho h$ as $j = u^t u_\varphi$ and write
\begin{equation}
\label{bernoulli3}
\int j(\Omega) d\Omega + \ln \left( \frac{h}{u^t} \right) + \int \frac{d(b^2 \mathcal L)}{2 \rho h \mathcal L} = C.
\end{equation}
This stays in agreement with the purely hydrodynamical case, where a functional relation $j =j(\Omega)$ (the rotation law) is an integrability condition of the Euler equations. Here we assume that $j = j(\Omega)$. It follows that also $b^2 \mathcal L$ must be a function of $\rho h \mathcal L$. Further details of the Euler-Bernoulli equation, as well as the specific choices regarding the equation of state, the rotation law, and the prescription of the magnetic field will be discussed in Sec.\ \ref{details_euler}.

\subsection{Einstein equations}

Following \cite{shibata} we derive the set of equations corresponding to a stationary black hole--torus spacetime from the standard $3+1$ formulation of the Einstein equations. The 3+1 metric reads
\begin{equation}
\label{form3p1}
g = (-\alpha^2 + \beta_i \beta^i) dt^2 + 2 \beta_i dx^i dt + \gamma_{ij} dx^i dx^j,
\end{equation}
where $\alpha$ is the lapse function, $\beta_i$ is the shift vector, and $\gamma_{ij}$ denote the components of the spatial metric. The vector normal to a surface of constant time $\Sigma_t$ is given by
\begin{equation}
n^\mu = \frac{1}{\alpha}(1,-\beta^r, -\beta^\theta, -\beta^\varphi), \quad n_\mu = (-\alpha,0,0,0).
\end{equation}

The Einstein constraint equations read
\begin{equation}
\label{momentum_constraint}
D_j (K^{ij} - \gamma^{ij} K) = 8 \pi j^i
\end{equation}
and
\begin{equation}
\label{hamiltonian_constraint}
\frac{1}{2} \left( R + K^2 - K_{ij}K^{ij} \right) = 8 \pi \rho_\mathrm{H},
\end{equation}
where $D_i$ and $R$ denote, respectively, the covariant derivative and the scalar curvature with respect to the metric $\gamma_{ij}$, induced on the slices $\Sigma_t$. The extrinsic curvature $K_{ij}$ of a slice $\Sigma_t$ is defined by
\begin{equation}
K_{ij} = -\frac{1}{2 \alpha} \left( \partial_t \gamma_{ij} - \mathcal L_\beta \gamma_{ij} \right),
\end{equation}
where the Lie derivative of the three-metric is given by
\begin{equation}
\mathcal L_\beta \gamma_{ij} = \beta^k \partial_k \gamma_{ij} + \gamma_{ik} \partial_j \beta^k + \gamma_{kj} \partial_i \beta^k.
\end{equation}
We denote $K = \gamma^{ij}K_{ij}$ and use a standard convention that spatial indices are raised and lowered using the induced metric $\gamma_{ij}$. The source terms $\rho_\mathrm{H}$ and $j^i$ are defined as
\begin{equation}
\rho_\mathrm{H} = n^\mu n^\nu T_{\mu \nu}, \quad j_i = - P\indices{^\alpha_i} n^\beta T_{\alpha \beta},
\end{equation}
where $P\indices{^\mu_\nu} = \delta\indices{^\mu_\nu} + n^\mu n_\nu$ is the spatial projection operator. The evolution equation for the extrinsic curvature $K_{ij}$ is
\begin{eqnarray}
\nonumber
\partial_t K_{ij} - \mathcal L_\beta K_{ij} & = & - D_i D_j \alpha + \alpha \left( R_{ij} + K K_{ij} - 2 K_{ik}K\indices{^k_j} \right) \\
& & + 4 \pi \alpha \left[ \gamma_{ij} (S - \rho_\mathrm{H}) - 2 S_{ij} \right].
\label{evolution_kij}
\end{eqnarray}
Here $R_{ij}$ is the Ricci tensor with respect to the metric $\gamma_{ij}$. The tensor $S_{ij}$ is defined as
\begin{equation}
S_{ij} = P\indices{^\mu_i} P\indices{^\nu_j} T_{\mu \nu},
\end{equation}
and $S = \gamma^{ij}S_{ij}$.

We start by computing the source terms $\rho_\mathrm{H}$, $j_i$, $S_{ij}$ and $S$. Assuming the energy momentum-tensor (\ref{emtensor}), one gets
\begin{eqnarray}
\rho_\mathrm{H} & = & \alpha^2 \rho h (u^t)^2 - p + \frac{1}{2} b^2, \\
S_{ij} & = & (\rho h + b^2)u_i u_j + \left( p + \frac{1}{2} b^2 \right) \gamma_{ij} - b_i b_j, \\
S & = & \rho h \left[ \alpha^2 (u^t)^2 - 1 \right] + 3 p + \frac{1}{2} b^2.
\end{eqnarray}
Finally, the only nonvanishing component of $j_i$ is
\begin{equation}
j_\varphi = \alpha \rho h u^t u_\varphi.
\end{equation}
Note that there is no explicit magnetic contribution to $j_\varphi$. All these formulas can be obtained quite generally, assuming the metric of the form (\ref{generalmetric}) and the conditions $u^r = u^\theta = b^r = b^\theta = 0$.

As in~\cite{shibata} we assume from now on a metric in quasi-isotropic form:
\begin{eqnarray}
\nonumber
g & = & - \alpha^2 dt^2 + \psi^4 e^{2q} (dr^2 + r^2 d\theta^2) + \\
&& \psi^4 r^2 \sin^2 \theta (\beta dt + d \varphi)^2.
\label{isotropic}
\end{eqnarray}
Thus, we need to provide equations for the four metric potentials appearing in Eq.~(\ref{isotropic}), $\alpha$, $\beta$, $\psi$, and $q$, or, as we shall see, suitable combinations of these quantities (as in~\cite{shibata}). In Eq.~(\ref{isotropic}) $\beta_\varphi = \psi^4 r^2 \beta \sin^2 \theta $, $\beta\equiv\beta^\varphi$, and $\alpha$ denotes the lapse function [as in Eq.\ (\ref{form3p1})]. Since $\partial_t \gamma_{ij} = 0$ and $\beta^k \partial_k \gamma_{ij} = 0$, we get
\begin{equation}
K_{ij} = \frac{1}{2 \alpha} \left( \gamma_{ik} \partial_j \beta^k + \gamma_{kj} \partial_i \beta^k \right),
\end{equation}
and $K = \gamma^{ik} K_{ik} = 0$, i.e., we are in fact working in a maximal slicing. Therefore, the momentum constraint (\ref{momentum_constraint}) can be written as
\begin{equation}
\label{momentum_constraint_2}
D_j K\indices{^j_l} = 8 \pi j_l.
\end{equation}
The only nonvanishing components of $K_{ij}$ read
\begin{eqnarray}
K_{\varphi r} &=& K_{r \varphi} = \frac{1}{2 \alpha} \psi^4 r^2 \sin^2 \theta \partial_r \beta, \\
K_{\varphi \theta} &=& K_{\theta \varphi} = \frac{1}{2 \alpha} \psi^4 r^2 \sin^2 \theta \partial_\theta \beta. 
\end{eqnarray}
To compute the momentum constraint we use a standard formula for symmetric tensors $K_{ij}$:
\begin{equation}
D_j K\indices{^j_l} = \frac{1}{\sqrt{\gamma}} \partial_j \left( \sqrt{\gamma} K\indices{^j_l} \right) - \frac{1}{2} (\partial_l \gamma_{ik}) K^{ik},
\end{equation}
where $\gamma=\det({\gamma_{ij}})$. Assuming the metric of the form (\ref{isotropic}) one obtains $\sqrt{\gamma} = \psi^6 r^2 \sin \theta e^{2 q}$. The only nontrivial component of Eqs.\ (\ref{momentum_constraint_2}) is obtained for $l = \varphi$,
\begin{eqnarray}
\frac{1}{r^2} \partial_r \left( \psi^2 r^2 K_{r \varphi} \right) + && \nonumber \\
\frac{1}{r^2 \sin \theta} \partial_\theta \left( \psi^2 \sin \theta K_{\theta \varphi} \right) & = & 8 \pi \psi^6 e^{2 q} j_\varphi.
\label{mom-cons-3}
\end{eqnarray}

Following~\cite{shibata} the shift vector $\beta$ can be now split into two parts, $\beta = \beta_\mathrm{K} + \beta_\mathrm{T}$, where subindex K indicates the Kerr metric and subindex T the torus. We require that
\begin{equation}
\label{Krf}
K_{r \varphi} = K_{\varphi r} = \frac{H_\mathrm{E} \sin^2 \theta}{\psi^2 r^2} +  \frac{1}{2 \alpha} \psi^4 r^2 \sin^2 \theta \partial_r \beta_\mathrm{T},
\end{equation}
\begin{equation}
\label{Ktf}
K_{\theta \varphi} = K_{\varphi \theta} = \frac{H_\mathrm{F} \sin \theta}{\psi^2 r} + \frac{1}{2 \alpha} \psi^4 r^2 \sin^2 \theta \partial_\theta \beta_\mathrm{T},
\end{equation}
and assume $H_\mathrm{E}$ and $H_\mathrm{F}$ corresponding to the Kerr metric (see Appendix A). More precisely, we choose $H_\mathrm{E}$ and $H_\mathrm{F}$ so that for the Kerr metric written in the form (\ref{isotropic}) one has
\begin{eqnarray}
K_{r \varphi} &=& K_{\varphi r} = \frac{H_\mathrm{E} \sin^2 \theta}{\psi^2 r^2}, \\
K_{\theta \varphi} &=& K_{\varphi \theta} = \frac{H_\mathrm{F} \sin \theta}{\psi^2 r}. 
\end{eqnarray}
These functions satisfy the momentum constraint of the form
\begin{eqnarray}
r \sin^3 \theta \partial_r H_\mathrm{E} + \partial_\theta (H_\mathrm{F} \sin^2 \theta) = 0. 
\end{eqnarray}

If a self-gravitating torus is present, we compute $\beta_\mathrm{K}$ from the relation
\begin{equation}
\label{betak1}
\partial_r \beta_\mathrm{K} = \frac{2 H_\mathrm{E} \alpha}{r^4 \psi^6},
\end{equation}
which does not yield the Kerr form, as the conformal factor $\psi$ contains a contribution from the torus.

Inserting the expressions for $K_{r \varphi}$ and $K_{\theta \varphi}$ into Eq.~(\ref{mom-cons-3}) we obtain, after some algebra, an elliptic-type equation for $\beta_\mathrm{T}$
\begin{eqnarray}
\Delta \beta_\mathrm{T} + \frac{\alpha}{\psi^6 r^2} (\partial_r \beta_\mathrm{T}) \partial_r \left( \frac{\psi^6 r^2}{\alpha} \right) + && \nonumber  \\
 \frac{\alpha}{\psi^6 r^2 \sin^2 \theta} (\partial_\theta \beta_\mathrm{T}) \partial_\theta \left( \frac{\psi^6 \sin^2 \theta}{\alpha} \right) & = & \frac{16 \pi \alpha e^{2q} j_\varphi}{r^2 \sin^2 \theta}, \nonumber \\
\end{eqnarray}
where $\Delta$ denotes the flat Laplacian operator in spherical coordinates. Again, as in~\cite{shibata} we replace the lapse function $\alpha$ by the combination $\Phi = \alpha \psi$ and rewrite the previous equation as
\begin{eqnarray}
\Delta \beta_\mathrm{T} + \left( \frac{2}{r} + \frac{7 \partial_r \psi}{\psi} - \frac{\partial_r \Phi}{\Phi} \right) \partial_r \beta_\mathrm{T} + &&  \nonumber \\
\frac{1}{r^2} \left( 2 \cot \theta + \frac{7 \partial_\theta \psi}{\psi} - \frac{\partial_\theta \Phi}{\Phi} \right) \partial_\theta \beta_\mathrm{T} & = & \frac{16 \pi \alpha e^{2q} j_\varphi}{r^2 \sin^2 \theta}, \nonumber \\
\end{eqnarray}
which is the same as Eq.\ (20) of \cite{shibata}.

The equation for the conformal factor follows from the Hamiltonian constraint, Eq.~(\ref{hamiltonian_constraint}), which for metric~(\ref{isotropic}) reads
\begin{eqnarray}
 R - K_{ij}K^{ij} = 16 \pi \rho_\mathrm{H}. 
\end{eqnarray}
It can be easily shown that
\begin{eqnarray}
K_{ij}K^{ij} = \frac{2 A^2}{\psi^{12} e^{2q}}, 
\end{eqnarray}
where we use the short-hand notation
\begin{equation}
\label{a2formula}
A^2 = \frac{(\psi^2 K_{r \varphi})^2}{r^2 \sin^2 \theta} + \frac{(\psi^2 K_{\theta \varphi})^2}{r^4 \sin^2 \theta}.
\end{equation}
The Ricci scalar  can be computed as
\begin{eqnarray}
R = -\frac{8}{\psi^5 e^{2q}} \Delta \psi + \frac{1}{\psi^4} \tilde R, 
\end{eqnarray}
where
\begin{eqnarray}
\tilde R = - 2 e^{-2q} \left( \partial_{rr}q + \frac{1}{r} \partial_r q + \frac{1}{r^2} \partial_{\theta \theta} q \right). 
\end{eqnarray}
This allows us to write the Hamiltonian constraint (\ref{hamiltonian_constraint}) in the form (Eq.\ (19) of \cite{shibata})
\begin{equation}
\label{laplacian_psi}
\Delta \psi = \frac{1}{8} \psi e^{2q} \tilde R - \frac{1}{4} \frac{A^2}{\psi^7} - 2 \pi \psi^5 e^{2q} \rho_\mathrm{H}.
\end{equation}

The next equation follows from the general evolution equation for $K$. It can be obtained by computing the trace of Eq.\ (\ref{evolution_kij}):
\begin{eqnarray}
\partial_t K - \mathcal L_\beta K & = &  - \gamma^{ij} D_i D_j \alpha + \alpha (R + K^2) + \nonumber \\
&& 4 \pi \alpha (S - 3 \rho_\mathrm{H}).
\end{eqnarray}
Using the Hamiltonian constraint, this equation can also be written as
\begin{eqnarray}
 \partial_t K - \mathcal L_\beta K = - \gamma^{ij} D_i D_j \alpha  + \alpha K_{ij}K^{ij} + 4 \pi \alpha (S + \rho_\mathrm{H}). 
 \nonumber \\
\end{eqnarray}
Since
\begin{eqnarray}
 - \gamma^{ij} D_i D_j \alpha = - \frac{1}{\psi^5 e^{2q}} \Delta \Phi + \frac{\Phi}{\psi^6 e^{2q}} \Delta \psi,
 \end{eqnarray}
we obtain an elliptic-type equation for $\Phi$
\begin{eqnarray} 
\Delta \Phi = \frac{\Phi}{\psi} \Delta \psi + \frac{2 \Phi A^2}{\psi^8} + 4 \pi \Phi e^{2q} \psi^4 (\rho_\mathrm{H} + S),
\end{eqnarray}
or, in terms of $\tilde R$,
\begin{eqnarray} 
\Delta \Phi = \frac{1}{8} \Phi e^{2q} \tilde R + \frac{7 \Phi A^2}{4 \psi^8} + 2 \pi \Phi e^{2q} \psi^4 (2S + \rho_\mathrm{H}), 
\end{eqnarray}
which is Eq.\ (18) of \cite{shibata}.

The last equation (for the potential $q$) is obtained from Eq.\ (\ref{evolution_kij}). We define
\begin{eqnarray}
 I_{ij} & = & \partial_t K_{ij} = \mathcal L_\beta K_{ij} - D_i D_j \alpha + \nonumber \\
&& \alpha \left( R_{ij} + K K_{ij} - 2 K_{ik}K\indices{^k_j} \right) + \nonumber \\
&& 4 \pi \alpha \left[ \gamma_{ij} (S - \rho_\mathrm{H}) - 2 S_{ij} \right] = 0.
\end{eqnarray}
Consider the equation
\begin{equation}
I_{rr} + \frac{1}{r^2} I_{\theta \theta} - \frac{3 e^{2q}}{r^2 \sin^2 \theta} I_{\varphi \varphi} = 0.
\end{equation}
It yields, in particular, the term
\begin{eqnarray}
\lefteqn{- \frac{4 \pi}{r^2} \left( r^2 S_{rr} + S_{\theta \theta} - \frac{3 e^{2q}}{\sin^2 \theta} S_{\varphi \varphi} + e^{2q} r^2 \psi^4 S \right) = } \nonumber \\
&& - 8 \pi e^{2q} \psi^4 \left( p - \frac{\rho h u_\varphi^2}{\psi^4 r^2 \sin^2 \theta} + \frac{3}{2} b^2 \right),
\end{eqnarray}
and the equation
\begin{eqnarray}
\lefteqn{ \left( \partial_{rr} + \frac{1}{r} \partial_r + \frac{1}{r^2} \partial_{\theta \theta} \right) q = } \nonumber \\
& & - 8 \pi e^{2q} \psi^4 \left( p - \frac{\rho h u_\varphi^2}{\psi^4 r^2 \sin^2 \theta} + \frac{3}{2} b^2 \right) + \frac{3 A^2}{\psi^8} \nonumber \\
& & + 2 \left( \frac{1}{r} \partial_r + \frac{\cot \theta}{r^2} \partial_\theta \right) \ln (\Phi \psi) \nonumber \\
& & + \frac{4}{\Phi \psi} \left( \partial_r \Phi \partial_r \psi + \frac{1}{r^2} \partial_\theta \Phi \partial_\theta \psi \right).
\end{eqnarray}
This allows us to compute $\tilde R$. The result can be combined with Eq.\ (\ref{laplacian_psi}), yielding a new form of the elliptic equation for the conformal factor $\psi$
\begin{eqnarray}
\nonumber
\Delta \psi & = & - 2 \pi e^{2q} \psi^5 \left( \rho_\mathrm{H} - p - \frac{3}{2} b^2 + \frac{\rho h u_\varphi^2}{\psi^4 r^2 \sin^2 \theta} \right) - \frac{A^2}{\psi^7} \\
\nonumber && - \frac{1}{2} \psi \left( \frac{1}{r} \partial_r + \frac{\cot \theta}{r^2} \partial_\theta \right) \ln (\Phi \psi)\\
& & - \frac{1}{\Phi} \left( \partial_r \Phi \partial_r \psi + \frac{1}{r^2} \partial_\theta \Phi \partial_\theta \psi \right),
\end{eqnarray}
which corresponds to Eq.\ (30) of \cite{shibata}. A direct calculation then gives
\begin{equation}
\left( \Delta + \frac{1}{r} \partial_r + \frac{\cot \theta}{r^2} \partial_\theta \right) (\Phi \psi) = 16 \pi \Phi \psi^5 e^{2q} \left( p + \frac{1}{2} b^2 \right),
\end{equation}
which generalizes Eq.\ (31) of \cite{shibata}.

From the technical point of view, the black hole is introduced by specifying suitable boundary conditions. This can be done in an elegant manner by adapting the above equation to the ``puncture'' form (see~\cite{brandt:1997,krivan:1998}). Assuming that the puncture is located at $r=0$, we define
\begin{equation}
\label{puncture}
\psi = \left( 1 + \frac{r_\mathrm{s}}{r} \right) e^\phi, \quad \Phi = \left( 1 - \frac{r_\mathrm{s}}{r} \right) e^{-\phi} B,
\end{equation}
where $r_\mathrm{s} = \frac{1}{2}\sqrt{m^2 - a^2}$ is a radius of a coordinate sphere corresponding to the black hole horizon. Parameters $m$ and $a$ are chosen in such a way that in the Kerr spacetime with the asymptotic mass $m$ and the spin parameter $a$ the event horizon would be located at $r = r_\mathrm{s}$.

The above definitions lead to the following equations for the functions $\phi$, $B$, $\beta_\mathrm{T}$, and $q$:
\begin{widetext}
\begin{subequations}
\label{main_sys}
\begin{eqnarray}
\left[ \partial_{rr} + \frac{1}{r } \partial_r  + \frac{1}{r^2} \partial_{\theta \theta}  \right] q & = & S_q, \label{47}\\
\left[ \partial_{rr} + \frac{2 r  }{r^2 - r_\mathrm{s}^2} \partial_r + \frac{1}{r^2} \partial_{\theta \theta} + \frac{  \cot{\theta}}{r^2}  \partial_\theta \right] \phi & = & S_\phi, \label{44} \\
\left[ \partial_{rr} + \frac{3 r^2 +  r_\mathrm{s}^2}{r(r^2 - r_\mathrm{s}^2)} \partial_r + \frac{1}{r^2} \partial_{\theta \theta} + \frac{2 \cot{\theta}}{r^2}  \partial_\theta \right] B & = & S_B, \label{45} \\
\left[ \partial_{rr} + \frac{4 r^2 - 8 r_\mathrm{s} r + 2 r_\mathrm{s}^2}{r(r^2 - r_\mathrm{s}^2)} \partial_r + \frac{1}{r^2} \partial_{\theta \theta} + \frac{3 \cot{\theta}}{r^2}  \partial_\theta \right]  \beta_\mathrm{T} & = & S_{\beta_\mathrm{T}},  \label{46}
\end{eqnarray}
 \end{subequations}
where
\begin{subequations}
\label{sources}
\begin{eqnarray}
S_q & = & -8 \pi e^{2q} \left( \psi^4 p - \frac{\rho h u_\phi^2}{r^2 \sin^2 \theta} + \frac{3}{2} \psi^4 b^2 \right) + \frac{3 A^2}{\psi^8} + 2 \left[ \frac{r - r_\mathrm{s}}{r(r + r_\mathrm{s})} \partial_r + \frac{\cot \theta}{r^2} \partial_\theta \right] \tilde b \\
\nonumber & & + \left[ \frac{8 r_\mathrm{s}}{r^2 - r_\mathrm{s}^2} + 4 \partial_r (\tilde b - \phi) \right] \partial_r \phi + \frac{4}{r^2} \partial_\theta \phi \partial_\theta (\tilde b - \phi), \\
S_\phi & = & - 2 \pi e^{2q} \psi^4 \left[ \rho_\mathrm{H} - p + \frac{\rho h u_\phi^2}{\psi^4 r^2 \sin^2 \theta} - \frac{3}{2} b^2 \right] - \frac{A^2}{\psi^8} \\
\nonumber & & - \partial_r\phi \partial_r \tilde b - \frac{1}{r^2} \partial_\theta \phi \partial_\theta \tilde b - \frac{1}{2} \left[ \frac{r - r_\mathrm{s}}{r (r + r_\mathrm{s})} \partial_r \tilde b + \frac{\cot \theta}{r^2} \partial_\theta \tilde b \right], \\
S_B & = & 16 \pi B e^{2q} \psi^4 \left( p + \frac{1}{2} b^2 \right), \\
S_{\beta_\mathrm{T}} & = & \frac{16 \pi \alpha e^{2q} j_\varphi}{r^2 \sin^2 \theta} - 8 \partial_r \phi \partial_r \beta_\mathrm{T} + \partial_r \tilde b \partial_r \beta_\mathrm{T} - 8 \frac{\partial_\theta \phi \partial_\theta \beta_\mathrm{T}}{r^2} + \frac{\partial_\theta \tilde b \partial_\theta \beta_\mathrm{T}}{r^2},
\end{eqnarray}
\end{subequations}
\end{widetext}
which replace Eqs.\ (44--47) of \cite{shibata} when a toroidal magnetic field is present in the disk. In the above formulas we denoted $B = e^{\tilde b}$. Equation (\ref{betak1}) can be written as
\begin{equation}
\label{betak_eq}
\partial_r \beta_\mathrm{K} = 2 H_\mathrm{E} B e^{-8 \phi} \frac{(r - r_\mathrm{s})r^2}{(r + r_\mathrm{s})^7}.
\end{equation}
Notice that it does not yield the Kerr form, as there are contributions from the torus in both $B$ and $\phi$.

In our numerical approach we assume equatorial symmetry and solve equations (\ref{main_sys}), (\ref{sources}) and (\ref{betak_eq}) in the domain defined by $r \in (r_\mathrm{s},r_\infty)$, $\theta \in (0,\pi/2)$. Here $r_\infty$ is large, but finite.

The boundary conditions at $r = r_\mathrm{s}$ read
\begin{equation}
\partial_r q = \partial_r \phi = \partial_r B = \partial_r \beta_\mathrm{T} = 0.
\end{equation}
It can be shown that Eq.\ (\ref{46}) requires a more stringent condition, which we set as $\beta_\mathrm{T} = O[(r - r_\mathrm{s})^4]$, or equivalently, $\beta_\mathrm{T} = \partial_r \beta_\mathrm{T} = \partial_{rr} \beta_\mathrm{T} = \partial_{rrr} \beta_\mathrm{T} = 0$ at $r = r_\mathrm{s}$. In this choice, reflecting a freedom of fixing the splitting $\beta = \beta_\mathrm{T} + \beta_\mathrm{K}$, we follow \cite{shibata}; this choice has consequences in the definition of the angular momentum of the black hole (cf.\ Sec.\ \ref{masses_ang_mom}).

With the above boundary conditions, the two-surface $r = r_\mathrm{s}$ embedded in a hypersurface of constant time $\Sigma_t$ becomes a Marginally Outer Trapped Surface (MOTS) or the so-called apparent horizon. This can be easily demonstrated as follows. A MOTS is defined as a two-surface $\mathcal S$ embedded in $\Sigma_t$ on which the scalar expansion of the outgoing null geodesics
\begin{equation}
\theta_+ = H - K_{ij} m^i m^j + K
\end{equation}
vanishes. Here $H = D_i m^i$ denotes the mean curvature of the surface $\mathcal S$, and $m^i$ is a unit vector tangent to $\Sigma_t$ and normal to $\mathcal S$. For the two-surface $r = r_\mathrm{s}$, the components of the three-vector $m^i$ are given by $m^i = (m^r, m^\theta, m^\varphi) = (\psi^{-2}e^{-q},0,0)$. Consequently, at $r = r_\mathrm{s}$,
\begin{equation}
\theta_+ = H = \frac{1}{\psi^6 e^{2q} r^2} \partial_r \left( \psi^4 e^q r^2 \right),
\end{equation}
since both terms $K_{ij} m^i m^j$ and $K$ vanish. Using Eq.\ (\ref{puncture}), one can show that
\begin{equation}
\theta_+ = H = \frac{1}{4} e^{-2 \phi - q} \left( 4 \partial_r \phi + \partial_r q \right),
\end{equation}
at $r = r_\mathrm{s}$. It is now clear that the boundary conditions assumed at $r = r_\mathrm{s}$ imply that $\theta_+ = H = 0$. Note that the surface $r = r_\mathrm{s}$ is not only an apparent horizon, but it is also a minimal surface.

At the axis $\theta = 0$ we assume regularity conditions $\partial_\theta \phi = \partial_\theta B = \partial_\theta \beta_\mathrm{T} = 0$. Local flatness implies that $q = 0$ at $\theta = 0$. At the equator, we require symmetry conditions $\partial_\theta q = \partial_\theta \phi = \partial_\theta B = \partial_\theta \beta_\mathrm{T} = 0$.

The asymptotic expansions of $q$, $\phi$, $B$, and $\beta_\mathrm{T}$ are discussed in Sec.\ \ref{masses_ang_mom}. They are used to impose boundary conditions at $r = r_\infty$. Further details on the numerical implementation of the boundary and asymptotic conditions can be found in \cite{kkmmop2}.

\subsection{Details of the Euler-Bernoulli equation}
\label{details_euler}

We next discuss details of the Euler-Bernoulli equation, Eq.\ (\ref{bernoulli3}). The following three components have to be specified in order to obtain a solution: the equation of state, the rotation law $j = j(\Omega)$, and a prescription of the distribution of the magnetic field.

We assume the Keplerian rotation law derived in \cite{kkmmop, kkmmop2}, i.e.,
\begin{equation}
\label{keplerian_rl}
j(\Omega) = - \frac{1}{2} \frac{d}{d \Omega} \ln \left\{ 1 - \left[ a^2 \Omega^2 + 3 w^\frac{4}{3} \Omega^\frac{2}{3} (1 - a \Omega)^\frac{4}{3} \right] \right\}.
\end{equation}
This is an exact formula that characterizes the motion of circular geodesics at the equatorial plane of the Kerr spacetime, in which case $w^2 = m$, where $m$ is the Kerr mass. For self-gravitating tori $w^2 \neq m$, in general. In the Newtonian limit rotation law (\ref{keplerian_rl}) yields the standard Keplerian prescription of the angular velocity $\Omega = w/(r \sin \theta)^\frac{3}{2}$. It also agrees (for $a = 0$) with the post-Newtonian Keplerian prescription proposed in \cite{mm2015}. The advantage of using this rotation law is that it allows one to obtain numerical solutions in a wide range of the parameters describing the torus \cite{kkmmop,kkmmop2}.

The angular velocity $\Omega$ can be obtained by solving the relation $j(\Omega) = u^tu_\varphi$ for $\Omega$. In more explicit terms this relation reads
\begin{eqnarray}
j(\Omega) \left( -g_{tt} - 2 g_{t\varphi} \Omega - g_{\varphi \varphi} \Omega^2 \right) = g_{\varphi \varphi} \Omega + g_{t \varphi} 
\end{eqnarray}
or
\begin{equation}
\label{rot_law_eq}
j(\Omega) \left[ \alpha^2 - \psi^4 r^2 \sin^2 \theta (\Omega + \beta)^2 \right] = \psi^4 r^2 \sin^2 \theta (\Omega + \beta),
\end{equation}
where $j(\Omega)$ is given by Eq.\ (\ref{keplerian_rl}). We assume a convention with $\Omega > 0$. This can correspond both to a torus corotating with the black hole, if $a > 0$, or counterrotating, for $a < 0$.

The specification of the magnetic term
\begin{equation}
\int \frac{d (b^2 \mathcal L)}{2 \rho h \mathcal L} = \int \frac{d (b^2 |\mathcal L|)}{2 \rho h |\mathcal L|},
\end{equation}
with $\mathcal L = - \alpha^2 \psi^4 r^2 \sin^2 \theta$, is somewhat more arbitrary, in the sense that there seem to be no physical ``hints'' concerning its prescription. Assuming a functional relation of the form $b^2 |\mathcal L|  = f(x)$, where $x = \rho h | \mathcal L|$ (note that this functional relation fulfills the general relativistic version of the von Zeipel condition for a purely toroidal magnetic field \cite{Zanotti15}), we obtain
\begin{eqnarray} 
\int \frac{d (b^2 |\mathcal L|)}{2 \rho h |\mathcal L|} = \int \frac{f^\prime(x) dx}{2x}. 
\end{eqnarray}
Suppose that we would like to get
\begin{eqnarray}
 \int \frac{f^\prime(x) dx}{2x} = \ln (1 + C_1 x)^n, 
\end{eqnarray}
where $C_1$ and $n$ are constants. This yields
\begin{eqnarray} 
f^\prime(x) = \frac{2 n C_1 x}{1 + C_1 x}, 
\end{eqnarray}
and a solution of the form
\begin{eqnarray} 
f(x) = 2 n \left[ x - \frac{1}{C_1} \ln (1 + C_1 x) \right] + C_2. 
\end{eqnarray}
For $x = 0$ we get $f(x = 0) = C_2$. Consequently, we set $C_2 = 0$. This ensures that the magnetic field vanishes for vanishing $\rho$.  We have, finally
\begin{eqnarray} 
\int \frac{d (b^2 \mathcal L)}{2 \rho h \mathcal L} = \ln \left[ \left( 1 + C_1 \alpha^2 \psi^4 r^2 \sin^2 \theta \rho h \right)^n \right] . 
\end{eqnarray}
We assume the above prescription of the magnetic field in this paper.

Equation (\ref{ut}) with the metric terms of Eq.~(\ref{isotropic}) yields
\begin{equation}
\frac{1}{u^t} = \sqrt{ \alpha^2 - \psi^4 r^2 \sin^2 \theta (\Omega + \beta)^2}.
\end{equation}
The Euler-Bernoulli Eq.\ (\ref{bernoulli3}) can be now written in the form
\begin{eqnarray}
\label{bernoulli2}
\lefteqn{h \left( 1 + C_1 \alpha^2 \psi^4 r^2 \sin^2 \theta \rho h \right)^n } \nonumber \\
&& \times \sqrt{ \alpha^2 - \psi^4 r^2 \sin^2 \theta (\Omega + \beta)^2} \nonumber \\
&& \times \left\{ 1 - \left[ a^2 \Omega^2 + 3 w^\frac{4}{3} \Omega^\frac{2}{3} (1 - a \Omega)^\frac{4}{3} \right] \right\}^{-\frac{1}{2}} = C^\prime.
\end{eqnarray}

In this paper we work with the polytropic equation of state of the form $p = K\rho^\gamma$. This yields the expression for the specific enthalpy
\begin{eqnarray}
\label{polytropic_h}
h = 1 + \frac{K \gamma}{\gamma - 1} \rho^{\gamma - 1}.
\end{eqnarray}

Note that the magnetic distribution was chosen in such a way that in the limit $\rho \to 0$, the Euler-Bernoulli equation has the same form as in the absence of the magnetic field, i.e.,
\begin{equation}
\label{eq_r1r2}
\frac{ \sqrt{ \alpha^2 - \psi^4 r^2 \sin^2 \theta (\Omega + \beta)^2} }{ \sqrt{ 1 - \left[ a^2 \Omega^2 + 3 w^\frac{4}{3} \Omega^\frac{2}{3} (1 - a \Omega)^\frac{4}{3} \right] } } = C^\prime.
\end{equation}
In our numerical procedure this form is used to establish the constants $w$ and $C^\prime$, assuming that the torus is characterized by some fixed equatorial coordinate radii $r_1$ and $r_2$.

An important numerical aspect concerns the specification of the polytropic constant $K$ of the equation of state. It is adjusted during the numerical iterative procedure so that the maximum value of the density $\rho$ within the torus is fixed at an a priori prescribed value (see Sec.\ \ref{method} for an exact discussion of this point).

We note that another possibility of setting up the details of the Euler-Bernoulli equation is to assume the rotation law of the form $\tilde {j} = h u_\varphi = \tilde j (\Omega)$. This is, for instance, the choice used in \cite{shibata}. The Euler-Bernoulli equation is then given by Eq.~(A10) of \cite{shibata}. Such a formulation would suggest a different profile of $b^2$, given for instance by Eq.\ (A14) of \cite{shibata}. 

\section{Masses and angular momenta}
\label{masses_ang_mom}

The asymptotic Arnowitt-Deser-Misner mass can be computed as \cite{shibata}
\begin{eqnarray} 
M_\mathrm{ADM} = \sqrt{m^2 - a^2} + M_1, 
\end{eqnarray}
where
\begin{eqnarray} 
M_1 = - 2 \int_{r_\mathrm{s}}^\infty dr \int_0^{\pi/2} d \theta (r^2 - r_\mathrm{s}^2) \sin \theta S_\phi, 
\end{eqnarray}
and $m$ is the Kerr mass.
Defining the mass of the black hole is less straightforward. The central black hole is surrounded by a minimal two-surface located at $r = r_\mathrm{s}$ in the puncture method, on a fixed hypersurface of constant time. There is a collection of quantities that can be used to characterize the geometry of the horizon $r = r_\mathrm{s}$. The area of the horizon is given by
\begin{eqnarray}
A_\mathrm{H} = 4 \pi \int_0^{\pi/2} \psi^4 e^q r^2 \sin \theta d \theta, 
\end{eqnarray}
where the integral is evaluated at $r = r_\mathrm{s}$. One can also define (at $r = r_\mathrm{s}$)
$\Omega_\mathrm{H} = - \beta = -\beta_\mathrm{K} = \mathrm{const}$, 
and the surface gravity
$\kappa = \partial_r \alpha \psi^{-2} e^{-q} = \mathrm{const}$. It can be easily shown that $\kappa = B e^{-4 \phi - q}/8 r_\mathrm{s}$.
The angular momentum of the black hole can be defined as (see below)
\begin{eqnarray}
J_\mathrm{H} = \frac{1}{4} \int_0^{\pi/2} d \theta \left( \frac{r^4 \sin^3 \theta \psi^6 \partial_r \beta}{\alpha} \right)_{r=r_\mathrm{s}}. 
\end{eqnarray}
A mass defined at $r = r_\mathrm{s}$ as
\begin{eqnarray}
M_\mathrm{H} = \int_0^{\pi/2} \psi^2 r^2 \partial_r \alpha \sin \theta d \theta + 2 \Omega_\mathrm{H} J_\mathrm{H} ,
\end{eqnarray}
obeys the Smarr formula
\begin{eqnarray}
M_\mathrm{H} = \frac{\kappa}{4 \pi}A_\mathrm{H} + 2 \Omega_\mathrm{H} J_\mathrm{H}. 
\end{eqnarray}
The mass of the black hole used in this paper is defined differently. Following \cite{shibata} we adopt Christodoulou's formula \cite{christodoulou}. We define first the so-called irreducible mass
\begin{eqnarray}
M_\mathrm{irr} = \sqrt{\frac{A_\mathrm{H}}{16 \pi}}. 
\end{eqnarray}
Then the mass of the black hole is defined as
\begin{eqnarray}
 M_\mathrm{BH} = M_\mathrm{irr} \sqrt{1 + \frac{J_\mathrm{H}^2}{4 M_\mathrm{irr}^4}}. 
\end{eqnarray}

From the ADM mass and the black hole mass we can define the torus mass as $M_\mathrm{T} = M_\mathrm{ADM} - M_\mathrm{BH}$, as was done in~\cite{kkmmop,kkmmop2}. There is, however, another possibility for the mass measure of the torus:
\begin{eqnarray}
M_\mathrm{T} = 8 \pi \int_{r_\mathrm{s}}^\infty dr \int_0^{\pi/2} d\theta r^2 \sin \theta \alpha \psi^6 e^{2q} \left( -T\indices{^t_t} + \frac{1}{2}T\indices{^\mu_\mu} \right). 
\nonumber \\
\label{torus_mass}
\end{eqnarray}
It satisfies the relation
\begin{eqnarray} 
M_\mathrm{H} + M_\mathrm{T} = \sqrt{m^2 - a^2} + M_1 = M_\mathrm{ADM}. 
\end{eqnarray}
We use this relation as a test of the accuracy of our numerical solutions. A direct computation yields
\begin{eqnarray} 
-T\indices{^t_t} + \frac{1}{2}T\indices{^\mu_\mu} = - \rho h u^t u_t - \frac{1}{2}\rho h + p + \frac{1}{2}b^2. 
\end{eqnarray}
In the numerical code, we compute the above quantity as
\begin{eqnarray} 
-T\indices{^t_t} + \frac{1}{2}T\indices{^\mu_\mu} = -\frac{1}{2} \rho h + 2 p + \rho_\mathrm{H} - \beta \rho h u^t u_\varphi. 
\end{eqnarray}

Correspondingly, the angular momentum of the torus is defined as
\begin{eqnarray}
J_1 & = & \int \sqrt{- g} T\indices{^t_\varphi} d^3 x \nonumber \\
& = & 4 \pi \int_{r_\mathrm{s}}^\infty dr \int_0^{\pi/2} d\theta r^2 \sin \theta \alpha \psi^6 e^{2q} \rho h u^t u_\varphi.
\end{eqnarray}
This is a standard definition corresponding to the Killing vector $\eta^\mu = (0,0,0,1)$, and the conservation law $\eta^\nu \nabla_\mu T\indices{^\mu_\nu} = \nabla_\mu (T\indices{^\mu_\nu} \eta^\nu) = 0$ \cite{leshouches}. Note that $T\indices{^t_\varphi} = (\rho h + b^2)u^t u_\varphi - b^t b_\varphi = \rho h u^t u_\varphi$, i.e., the contributions from the magnetic terms cancel. The total, asymptotic angular momentum reads $J = J_\mathrm{H} + J_1$.

The value of the angular momentum $J_\mathrm{H}$ depends on the assumed boundary conditions for $\beta_\mathrm{T}$. In our case $\beta_\mathrm{T} = \partial_r \beta_\mathrm{T} = \partial_{rr} \beta_\mathrm{T} = \partial_{rrr} \beta_\mathrm{T} = 0$ at $r = r_\mathrm{s}$, and consequently $J_\mathrm{H} = a m$.

The mass $M_1$ and the angular momentum $J_1$ are related  to the asymptotic behaviour of the metric functions $\phi$ and $\beta_\mathrm{T}$, namely,
\begin{eqnarray} 
\phi \sim \frac{M_1}{2r}, \quad \beta_\mathrm{T} \sim -\frac{2 J_1}{r^3},  
\end{eqnarray}
as $r \to \infty$. The asymptotic behaviour of the two remaining functions $B$ and $q$ is given by
\begin{eqnarray}  
B \sim 1 - \frac{B_1}{r^2}, \quad  q \sim \frac{q_1 \sin^2 \theta}{r^2}, 
\end{eqnarray}
where
\begin{eqnarray}  
B_1 = \frac{2}{\pi} \int_{r_\mathrm{s}}^\infty dr \frac{(r^2 - r_\mathrm{s}^2)^2}{r} \int_0^{\pi/2} d \theta \sin^2 \theta S_B, 
\end{eqnarray}
and
\begin{eqnarray}
q_1 & = & \frac{2}{\pi} \int_{r_2}^\infty dr r^3 \int_0^{\pi/2} d\theta \cos(2 \theta) S_q \nonumber \\
&& - \frac{4 r_\mathrm{s}^2}{\pi} \int_0^{\pi/2} d\theta \cos(2 \theta) q(r_\mathrm{s}, \theta).
\end{eqnarray}
We use the above asymptotic expansions to set the boundary conditions at the outer boundary of the numerical grid, i.e., at $r = r_\infty$.

\begin{table*}
\caption{\label{table1}Parameters of the numerical solutions. In all cases we assumed the polytropic exponent $\gamma = 4/3$, the magnetisation law parameter $n = 1$, and the black hole mass parameter $m = 1$. From left to right the columns report: the black hole spin parameter $a$, the coordinate inner radius of the disk $r_1$, the circumferential inner radius of the disk $r_\mathrm{C,1}$, the coordinate outer radius of the disk $r_2$, the circumferential outer radius of the disk $r_\mathrm{C,2}$, the maximum rest-mass density within the disk $\rho_\mathrm{max}$, the parameter $C_1$ appearing in the magnetisation law, the total ADM mass $m_\mathrm{ADM}$, the mass of the black hole $m_\mathrm{BH}$, the angular momentum of the disk $J_1$, and the magnetisation parameter $\beta_\mathrm{mag}$.}
\begin{ruledtabular}
\begin{tabular}{ccccccccccccc}
No. & $a$ & $r_1$ & $r_\mathrm{C,1}$ & $r_2$ & $r_\mathrm{C,2}$ & $\rho_\mathrm{max}$ & $C_1$ & $m_\mathrm{ADM}$ & $m_\mathrm{BH}$ & $J_1$ & $\beta_\mathrm{mag}$ \\
\hline
1a & $-0.5$ & 8.0 & 9.2 & 35.3 & 36.7  & $5 \times 10^{-5}$ & 0      & 1.33 & 1.01 & 1.7   & $\infty$ \\
1b & $-0.5$ & 8.0 & 9.2 & 35.3 & 36.7  & $5 \times 10^{-5}$ & 0.01 & 1.34 & 1.01 & 1.75 & 30.5    \\
1c & $-0.5$ & 8.0 & 9.3 & 35.3 & 36.8  & $5 \times 10^{-5}$ & 0.1   & 1.40 & 1.01 & 2.08 & 3.49    \\
1d & $-0.5$ & 8.0 & 9.4 & 35.3 & 36.9  & $5 \times 10^{-5}$ & 1      & 1.51 & 1.02 & 2.65 & 0.21    \\
1e & $-0.5$ & 8.0 & 9.4 & 35.3 & 36.9  & $5 \times 10^{-5}$ & 1.3   & 1.50 & 1.02 & 2.58 & $5.3 \times 10^{-2}$ \\
1f & $-0.5$ & 8.0 & 9.3 & 35.3 & 36.9  & $5 \times 10^{-5}$ & 1.42  & 1.49 & 1.02 & 2.54 & $1.3 \times 10^{-3}$
\\
\hline
2a & 0 & 8.1 & 9.3 & 35.1 & 36.5    & $5 \times 10^{-5}$ & 0      &  1.33 & 1.02 & 1.64 & $\infty$ \\
2b & 0 & 8.1 & 9.3 & 35.1 & 36.5    & $5 \times 10^{-5}$ & 0.01 &  1.34 & 1.02 & 1.69 & 29.4    \\
2c & 0 & 8.1 & 9.3 & 35.1 & 36.5    & $5 \times 10^{-5}$ & 0.1   &  1.40 & 1.02 & 2.02 & 3.37    \\
2d & 0 & 8.1 & 9.4 & 35.1 & 36.7    & $5 \times 10^{-5}$ & 1      &  1.52 & 1.03 & 2.61 & 0.19    \\
2e & 0 & 8.1 & 9.4 & 35.1 & 36.7    & $5 \times 10^{-5}$ & 1.3   &  1.51 & 1.03 & 2.55 & $3.0 \times 10^{-2}$ \\
2f & 0 & 8.1 & 9.4 & 35.1 & 36.7    & $5 \times 10^{-5}$ & 1.37  &  1.50 & 1.03 & 2.52 & $5.8 \times 10^{-4}$
\\
\hline
3a & 0.9 & 3.0 & 4.4 & 20.0 & 21.7 &  $3.5 \times 10^{-4}$ & 0       &  1.52 & 1.00 &2.04& $\infty$ \\
3b & 0.9 & 3.0 & 4.4 & 20.0 & 21.7 &  $3.5 \times 10^{-4}$ & 0.01  &  1.52 & 1.00 &2.05& $75.8$ \\
3c & 0.9 & 3.0 & 4.4 & 20.0 & 21.7 &  $3.5 \times 10^{-4}$ & 0.1    &  1.55 & 1.00 &2.17& $8.38$ \\
3d & 0.9 & 3.0 & 4.4 & 20.0 & 21.7 &  $3.5 \times 10^{-4}$ & 1       &  1.57 & 1.01 &2.23& 0.96\\
3e & 0.9 & 3.0 & 4.4 & 20.0 & 21.6 &  $3.5 \times 10^{-4}$ & 2       &  1.47 & 1.00 &1.79& 0.26 \\
3f & 0.9 & 3.0 & 4.4 & 20.0 & 21.5 &  $3.5 \times 10^{-4}$ & 2.74   &  1.39 & 1.00 &1.45& $5.88 \times 10^{-4}$ \\
\hline
4a & 0.99 & 0.8 & 2.41 & 20.1 & 21.9 &  $1.5 \times 10^{-3}$ & 0       & 1.70 & 1.00 & 2.31 & $\infty$ \\
4b & 0.99 & 0.8 & 2.41 & 20.1 & 21.9 &  $1.5 \times 10^{-3}$ & 0.01  & 1.70 & 1.00 & 2.30 & 805.5    \\
4c & 0.99 & 0.8 & 2.40 & 20.1 & 21.9 &  $1.5 \times 10^{-3}$ & 0.1    & 1.68 & 1.00 & 2.24 & 80.3     \\
4d & 0.99 & 0.8 & 2.38 & 20.1 & 21.7 &  $1.5 \times 10^{-3}$ & 1       & 1.51 & 1.00 & 1.64 & 7.72     \\
4e & 0.99 & 0.8 & 2.35 & 20.1 & 21.5 &  $1.5 \times 10^{-3}$ & 2       & 1.32 & 1.00 & 1.01 & 3.07     \\
4f & 0.99 & 0.8 & 2.33 & 20.1 & 21.3 &  $1.5 \times 10^{-3}$ & 3       & 1.17 & 1.00 & 0.52 & 1.31     \\
4g & 0.99 & 0.8 & 2.32 & 20.1 & 21.3 &  $1.5 \times 10^{-3}$ & 4       & 1.08 & 1.00 & 0.24 & 0.39     \\
4h & 0.99 & 0.8 & 2.32 & 20.1 & 21.2 &  $1.5 \times 10^{-3}$ & 4.5    & 1.05 & 1.00 & 0.17 & 0.11     \\
4i & 0.99 & 0.8 & 2.32 & 20.1 & 21.2 &  $1.5 \times 10^{-3}$ & 4.7    & 1.05 & 1.00 & 0.15 & 2.28 $\times 10^{-2}$
\end{tabular}
\end{ruledtabular}
\end{table*}

\section{Numerical method}
\label{method}

To construct our models of self-gravitating, magnetised tori around rotating black holes we need to solve numerically the equations derived in Sec.~\ref{equations}. The metric functions are described by Eq.\ (\ref{betak_eq}) and Eqs.\ (\ref{main_sys}) with the source terms given by Eqs.\ (\ref{sources}). The angular velocity $\Omega$ must satisfy Eq.\ (\ref{rot_law_eq}) with $j(\Omega)$ given by Eq.\ (\ref{keplerian_rl}). The distribution of the enthalpy $h$, rest-mass density, and the pressure $p$ are obtained from Eq.\ (\ref{bernoulli2}) and from the polytropic relation (\ref{polytropic_h}). The quantity $A^2$ appearing in expressions (\ref{sources}) is defined by (\ref{a2formula}) where $K_{r \varphi}$ and $K_{\theta \varphi}$ should be computed according to formulas (\ref{Krf}) and (\ref{Ktf}).

The numerical code used to obtain the solutions presented in this paper is a modification of the code described and tested in \cite{kkmmop2} to which the interested reader is addressed for details. It is an iterative method, where in each Newton-Raphson  iteration one solves Eq.\ (\ref{rot_law_eq}) for the angular velocity $\Omega$, Eq.\ (\ref{bernoulli2}) for the density $\rho$ (or the specific enthalpy $h$), and then Eqs.\ (\ref{main_sys}) for the metric functions. The latter are solved with 2nd-order finite differences. We take advantage of the banded matrix structure of the resulting linear equations and use LAPACK \cite{lapack}. The changes introduced with respect to the version of the code described in \cite{kkmmop2} are only related to the presence of the magnetic field. While the inclusion of the magnetic terms in Eqs.\ (\ref{main_sys}) is straightforward,  solving Eq.\ (\ref{bernoulli2}) with the magnetic terms is more troublesome. To describe it we need to discuss details connected with the treatment of Eqs.\ (\ref{rot_law_eq}) and (\ref{bernoulli2}).

\begin{figure*}
\centering
\includegraphics[scale=0.17]{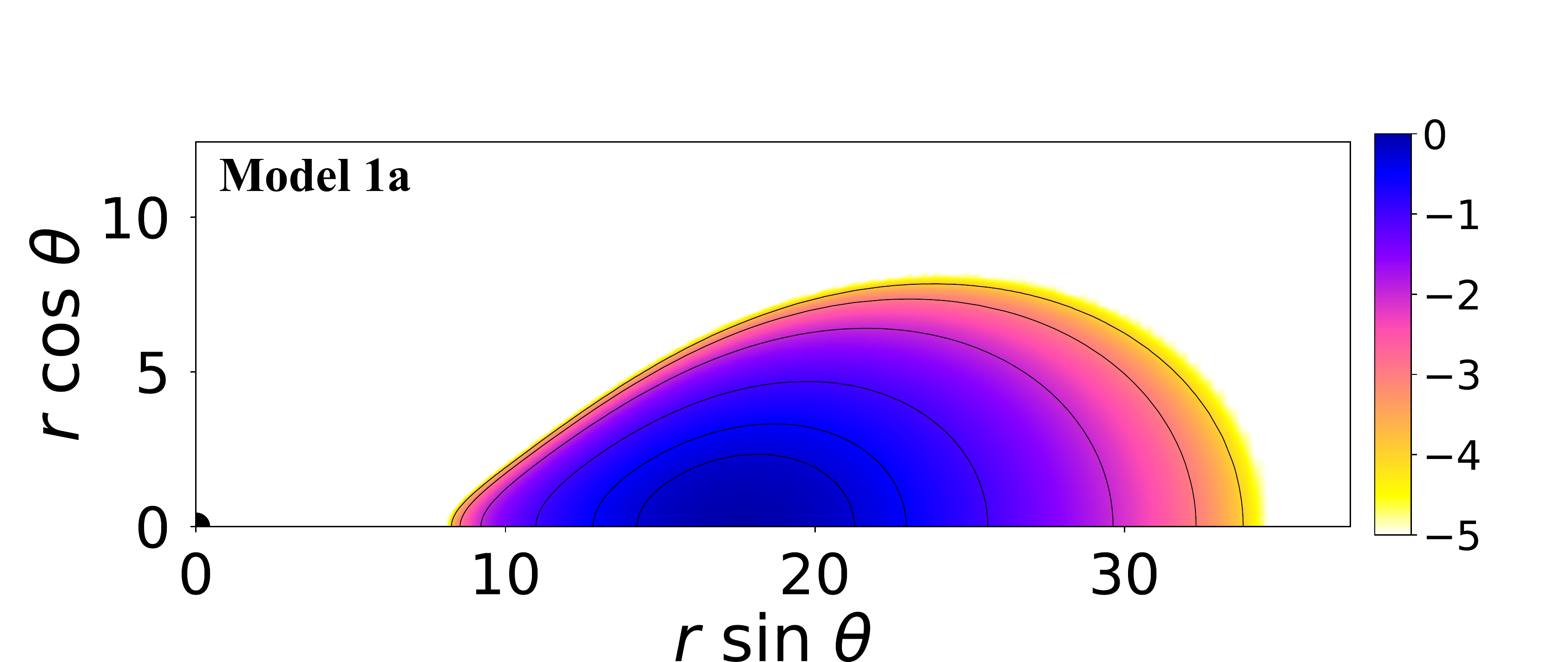}
\hspace{-0.4cm}
\includegraphics[scale=0.17]{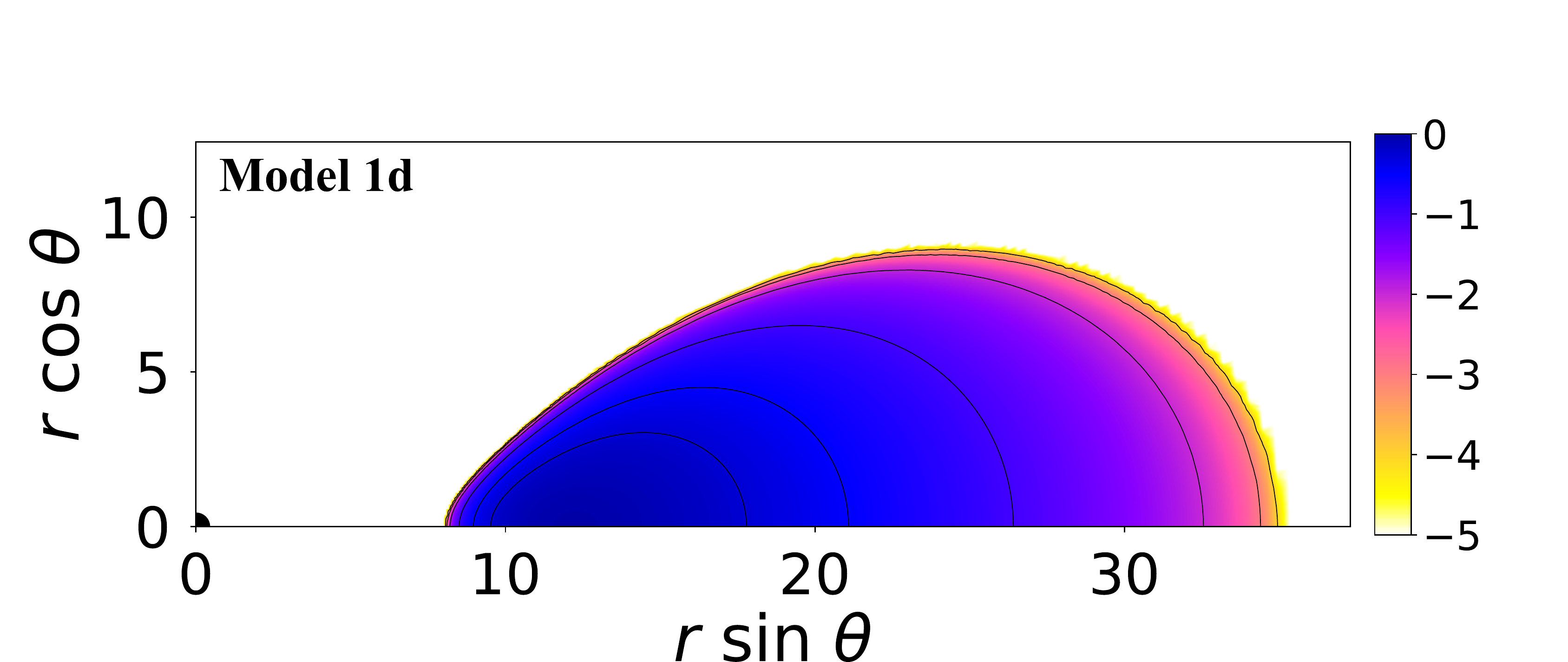}
\hspace{-0.4cm}
\includegraphics[scale=0.17]{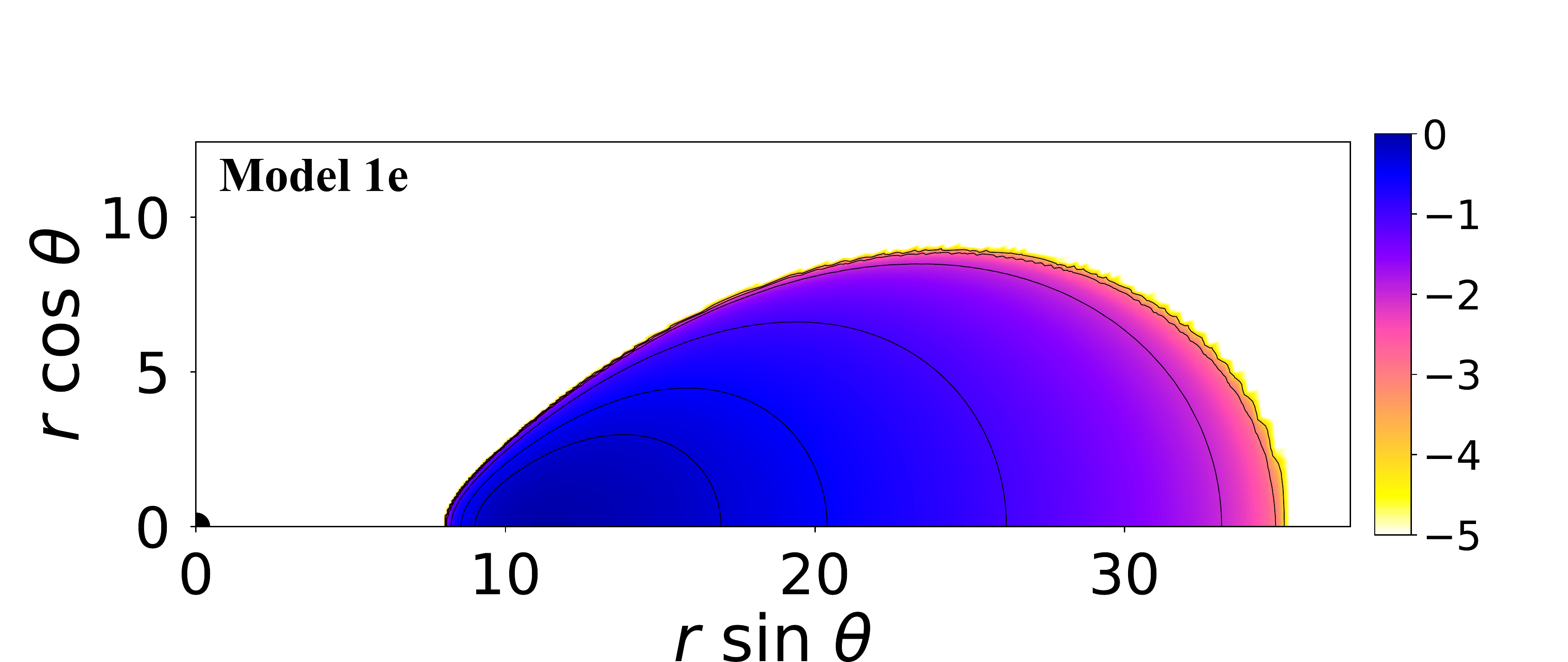}
\\
\includegraphics[scale=0.17]{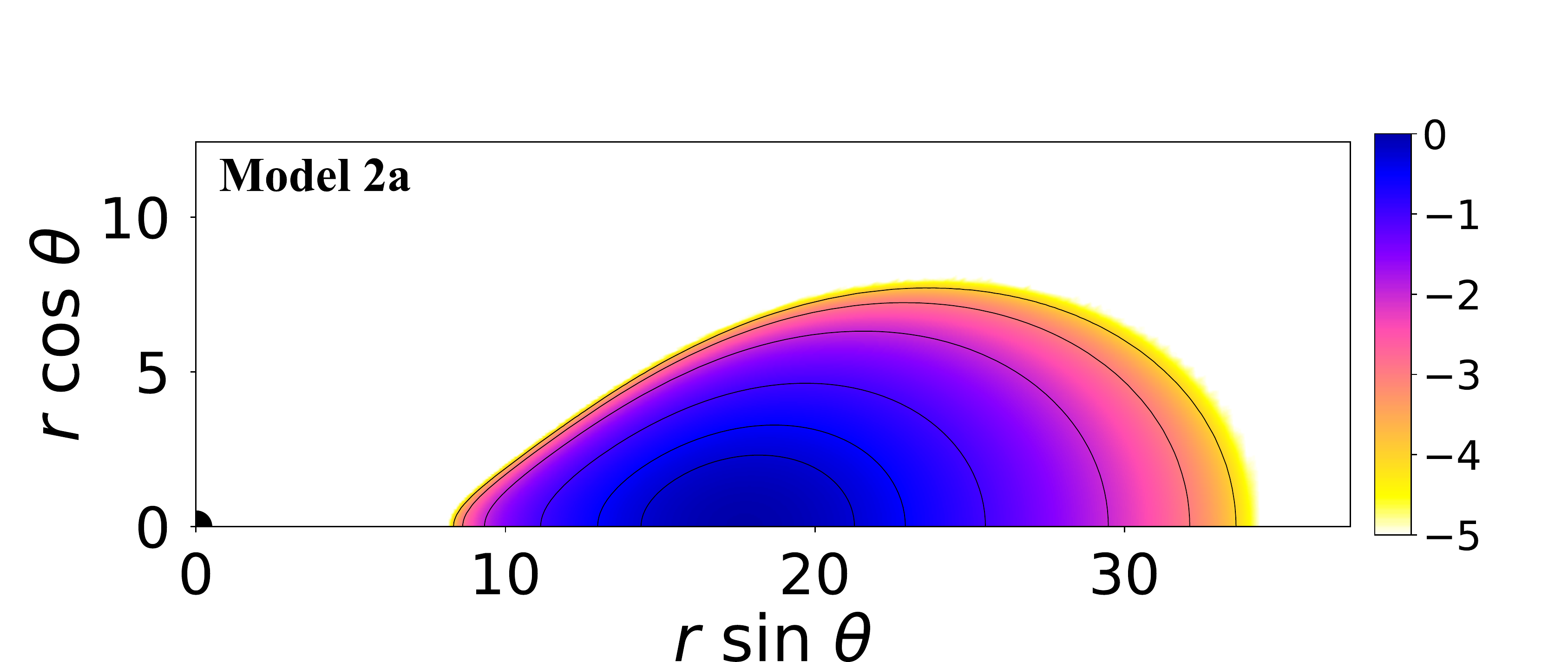}
\hspace{-0.4cm}
\includegraphics[scale=0.17]{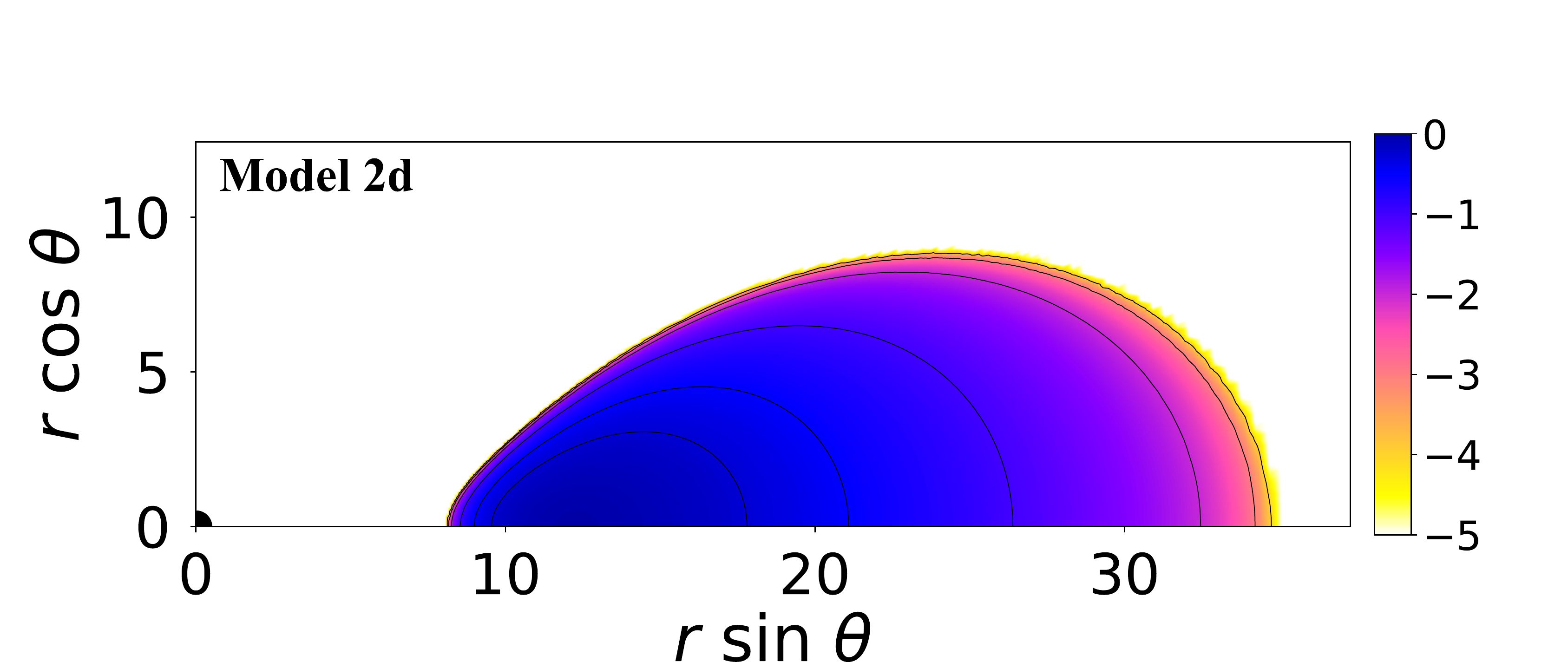}
\hspace{-0.4cm}
\includegraphics[scale=0.17]{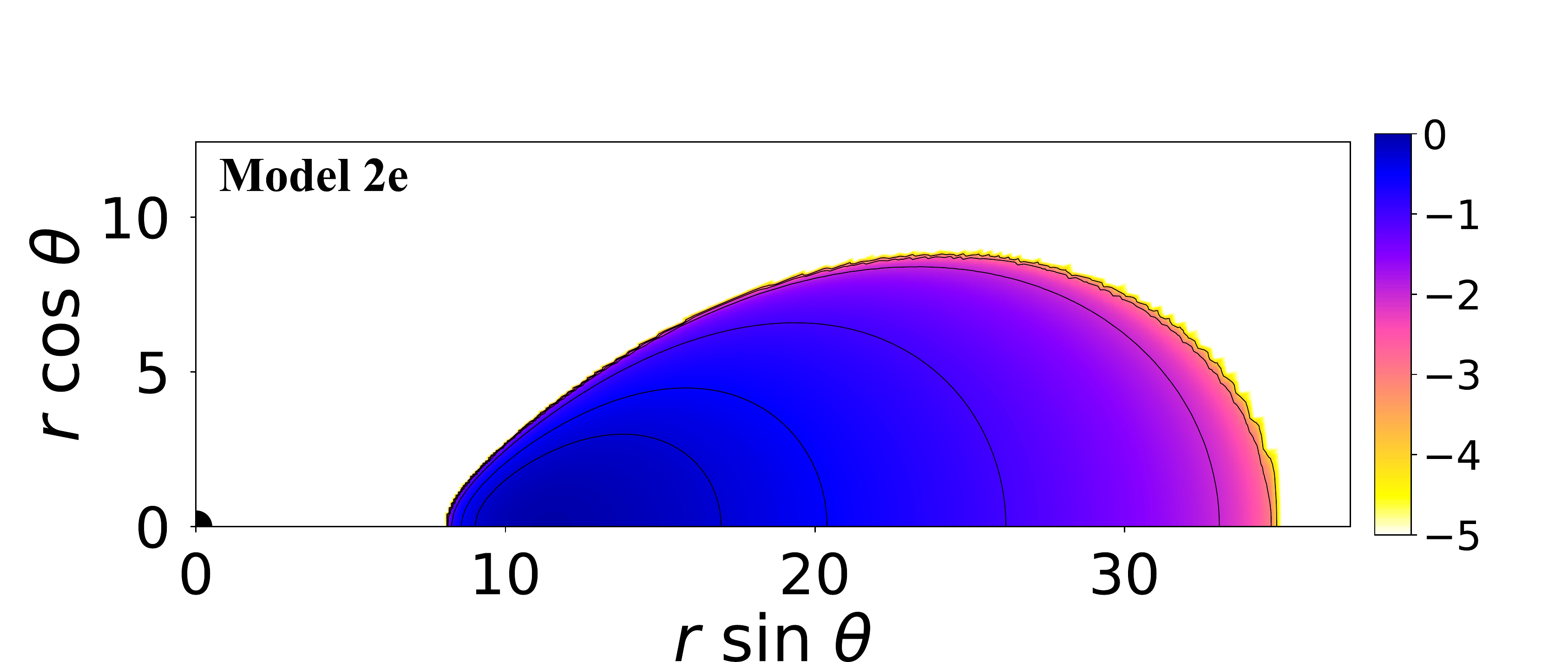}
\\
\includegraphics[scale=0.17]{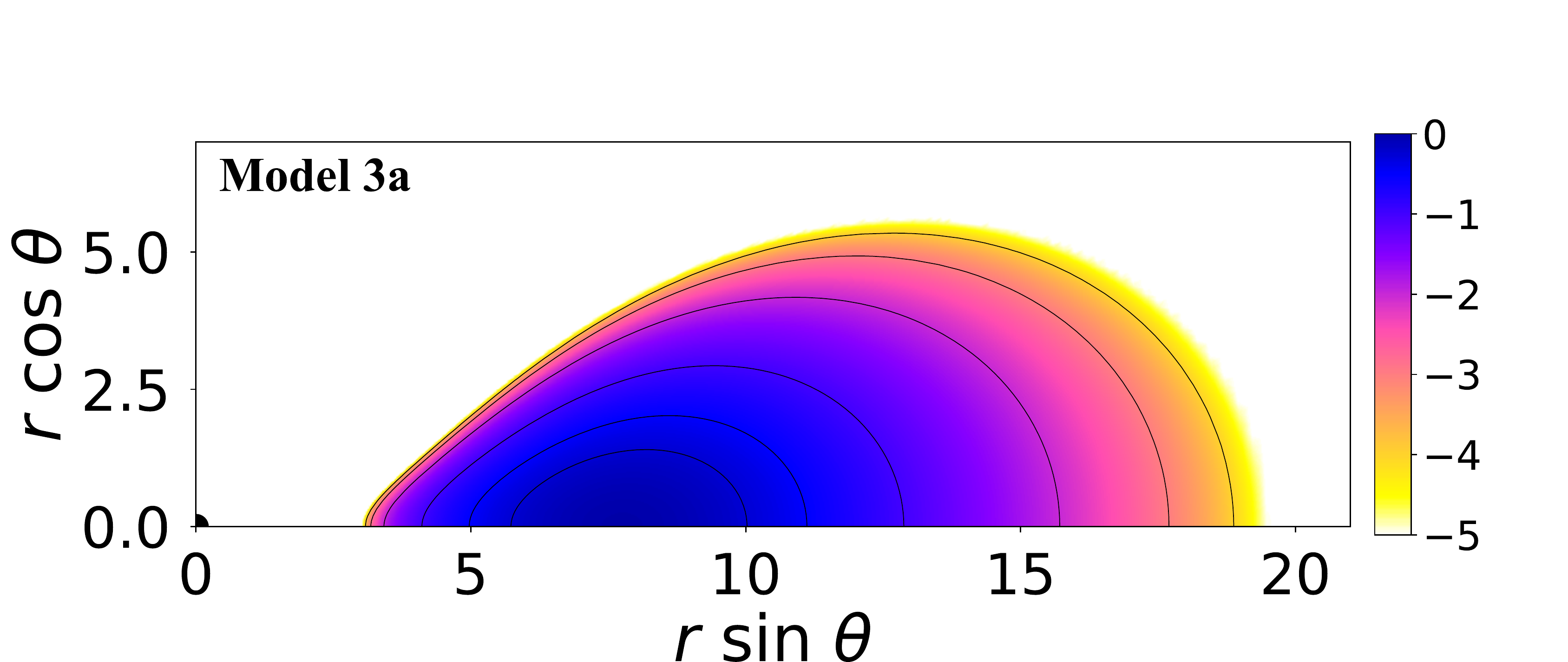}
\hspace{-0.4cm}
\includegraphics[scale=0.17]{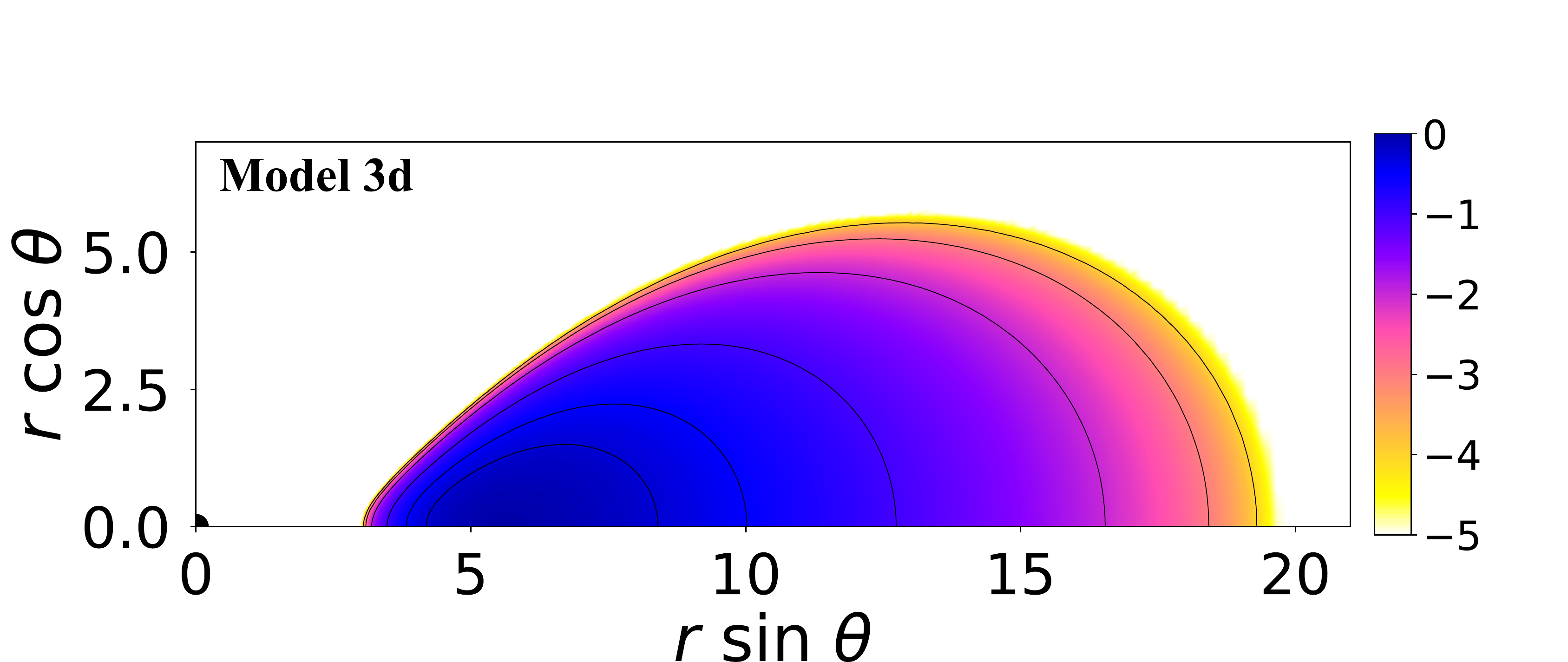}
\hspace{-0.4cm}
\includegraphics[scale=0.17]{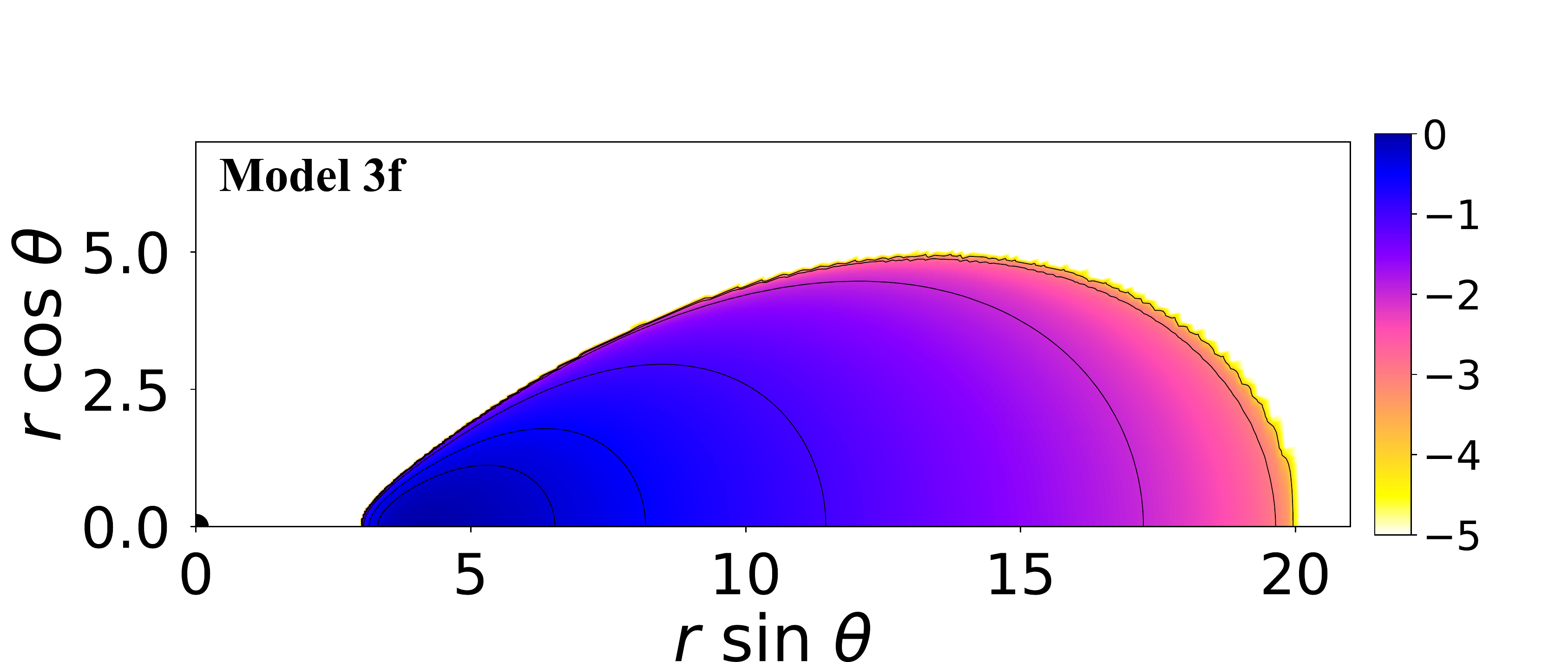}
\\
\includegraphics[scale=0.17]{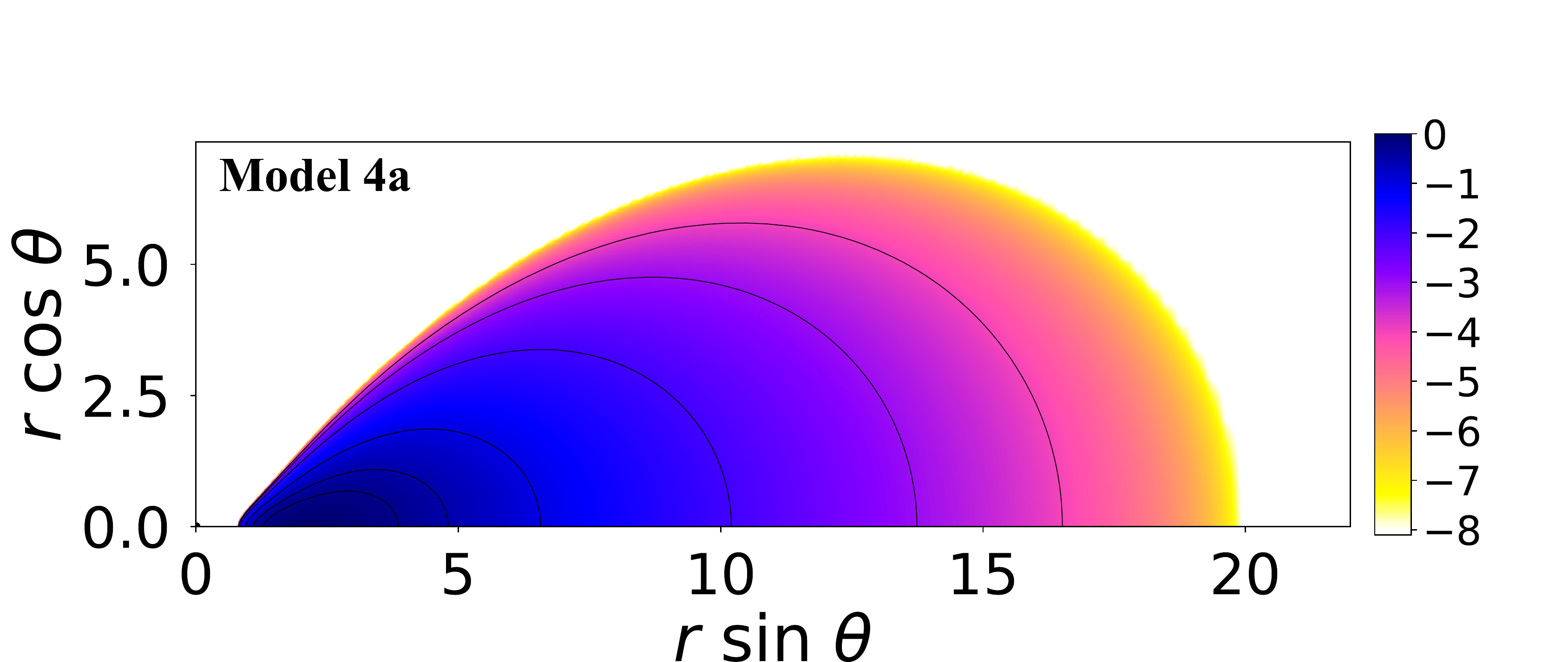}
\hspace{-0.4cm}
\includegraphics[scale=0.17]{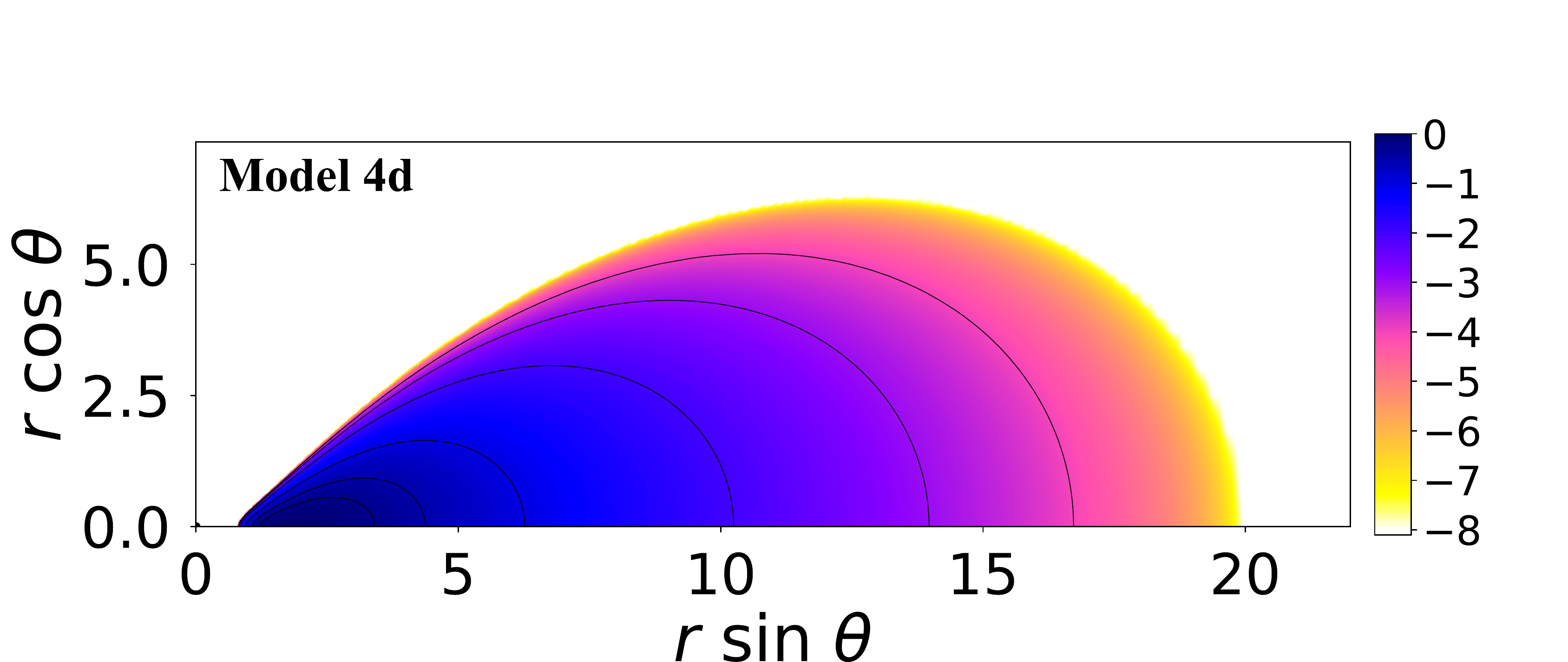}
\hspace{-0.4cm}
\includegraphics[scale=0.17]{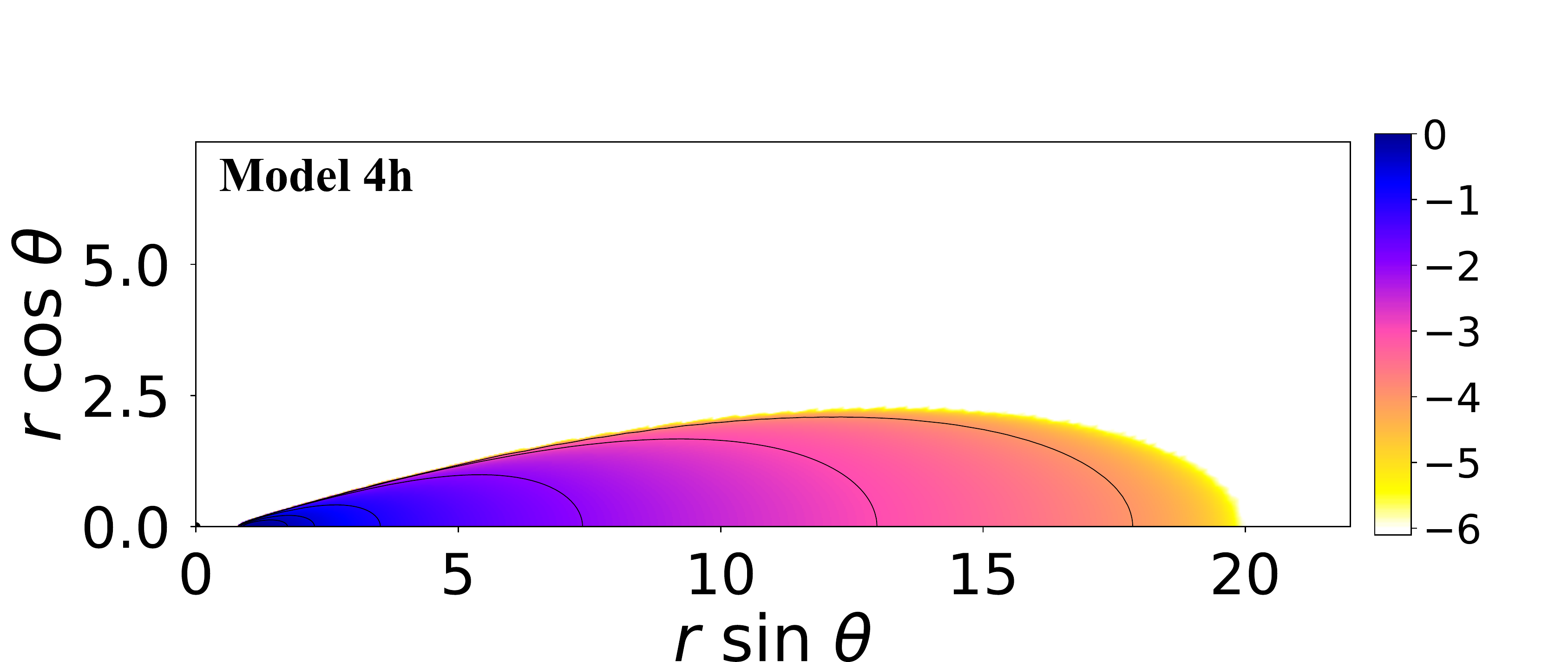}
\caption{Morphology of the disks: distribution of the logarithm of the rest-mass density for selected models of our sample (see Table~\ref{table1}). The effects of the magnetisation increase from left to right; the leftmost column depicts disks with no magnetic field.}
\label{models}
\end{figure*}

Each iteration is started with a Newton-Raphson procedure that gives the values of constants $w$ and $C^\prime$, assuming that the inner and outer equatorial radii of the disk ($r_1$ and $r_2$, respectively) are fixed. This procedure solves Eqs.\ (\ref{rot_law_eq}) and (\ref{eq_r1r2}) at points $(r,\theta) = (r_1,\pi/2), (r_2,\pi/2)$. These are four equations for the four unknowns $w$, $C^\prime$, $\Omega_1 = \Omega(r_1,\pi/2)$, $\Omega_2 = \Omega(r_2,\pi/2)$; in this step we assume that the metric functions are known from the previous iteration, or from the initial guess. In the next step we compute the values of $\Omega$ in a region which is large enough to contain the disk, but smaller than the domain covered by the numerical grid. In this way we can avoid problems with finding solutions to Eq.\ (\ref{rot_law_eq}) in the vicinity of the symmetry axis $\theta = 0$. Equation (\ref{rot_law_eq}) is also solved with a Newton-Raphson scheme. The next stage consists in solving Eq.\ (\ref{bernoulli2}) for the specific enthalpy $h$, also by a Newton-Raphson procedure. The problem that one encounters here (which is absent in the purely hydrodynamical case) is that Eq.\ (\ref{bernoulli2}) contains a density term $\rho$, and in order to obtain a solution for $h$ one has to specify the value of the polytropic constant $K$. On the other hand, in \cite{kkmmop2} we found that the possibility of obtaining a convergent solution increases considerably if the solution is parameterized by a maximum value of the rest-mass density $\rho_\mathrm{max}$ within the disk. In the purely hydrodynamical case the value of the polytropic constant is then adjusted at each iteration so that the maximum value of the specific enthalpy $h$ obtained from Eq.\ (\ref{bernoulli2}) (with no magnetic terms) corresponds to the maximum of $\rho$ equal to an a priori prescribed value $\rho_\mathrm{max}$. This approach is not straightforward in the present GRMHD case. Therefore, we instead take the value of the polytropic constant $K$ inherited from the previous iteration, solve Eq.\ (\ref{bernoulli2}) for $h$, and then assume a value of $K$ so that the maximum in the specific enthalpy $h$ corresponds to the maximum in $\rho$ equal to $\rho_\mathrm{max}$. This approach leads to convergent solutions.

All stationary solutions of self-gravitating, magnetised disks obtained in this work have been computed on a numerical grid with approximately 800 nodes in the radial direction and 200 nodes in the angular direction. Specifically, the nodes in the grid are distributed according to
\begin{equation}
r_i = r_\mathrm{s} + \frac{f^{i-1} - 1}{f - 1} \Delta r, \quad i = 1, \dots, N_r,
\end{equation}
in the radial direction, and
\begin{equation}
\theta_j =
\begin{cases}
0, & j = 1, \\
\arccos \left[ 1 + \left( \frac{3}{2} - j \right) \Delta \mu \right], & j = 2, N_\theta - 1, \\
\frac{\pi}{2}, & j = N_\theta,
\end{cases}
\end{equation}
where $\Delta \mu = 1/(N_\theta - 2)$, in the angular direction. We choose, in particular, $\Delta r = r_\mathrm{s}/50$, $f = 1.01$, $N_r = 800$, $N_\theta = 200$. The above grid specification is similar to the one used in \cite{shibata}. 

The number of iterations required to obtain a solution depends mainly on the resolution of the grid, but also on the parameters of the solution \cite{kkmmop2}. Obtaining the solutions collected in Table \ref{table1} required typically $\sim 10^4$ to $\sim 2 \times 10^4$ interations. Highly magnetised disks denoted in Table \ref{table1} as 4e--4i  are exceptional, and required up to $\sim 10^5$ iterations.

\section{Results}
\label{results}

\begin{figure*}
\centering
\includegraphics[width=0.9\columnwidth]{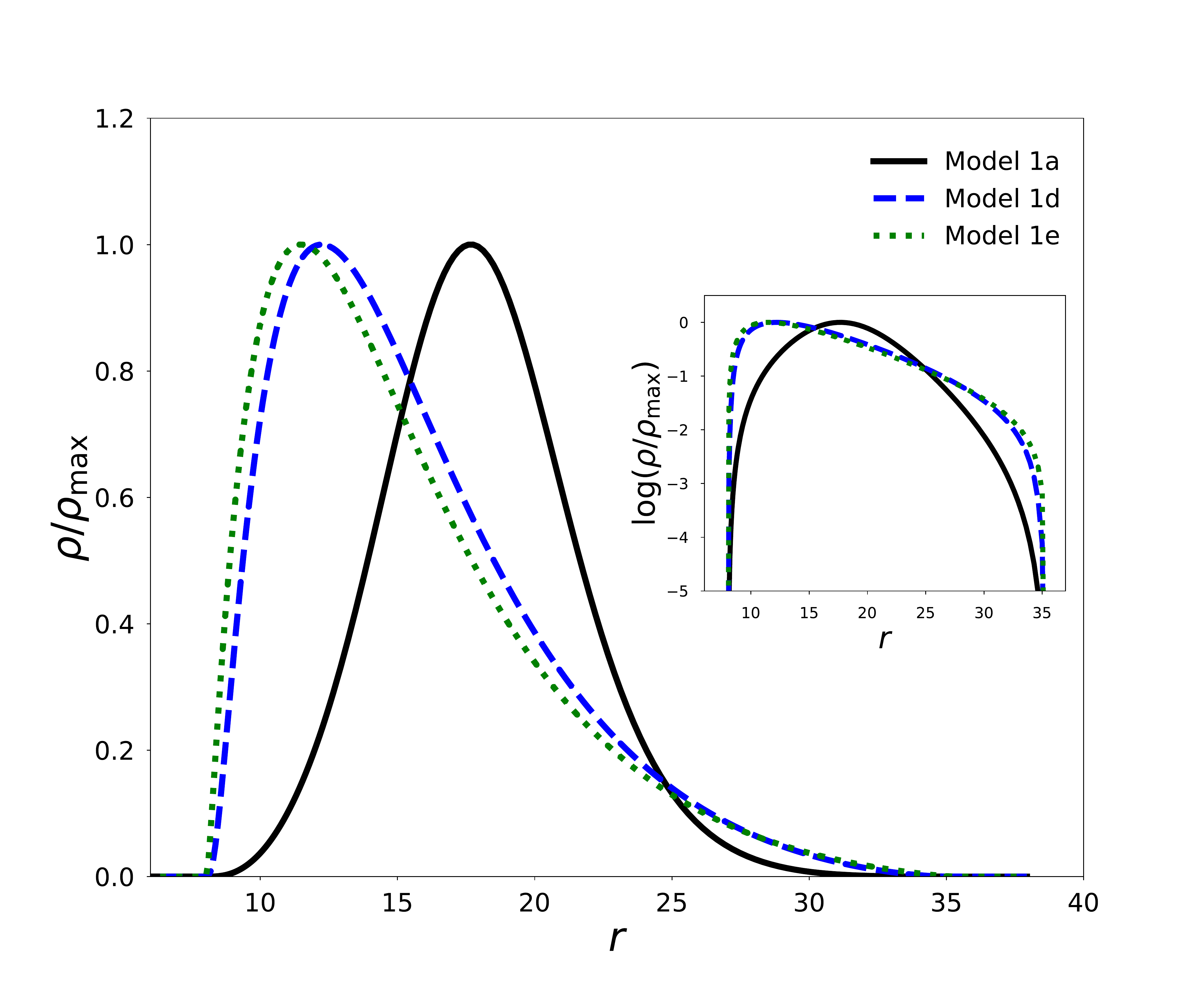}
\includegraphics[width=0.9\columnwidth]{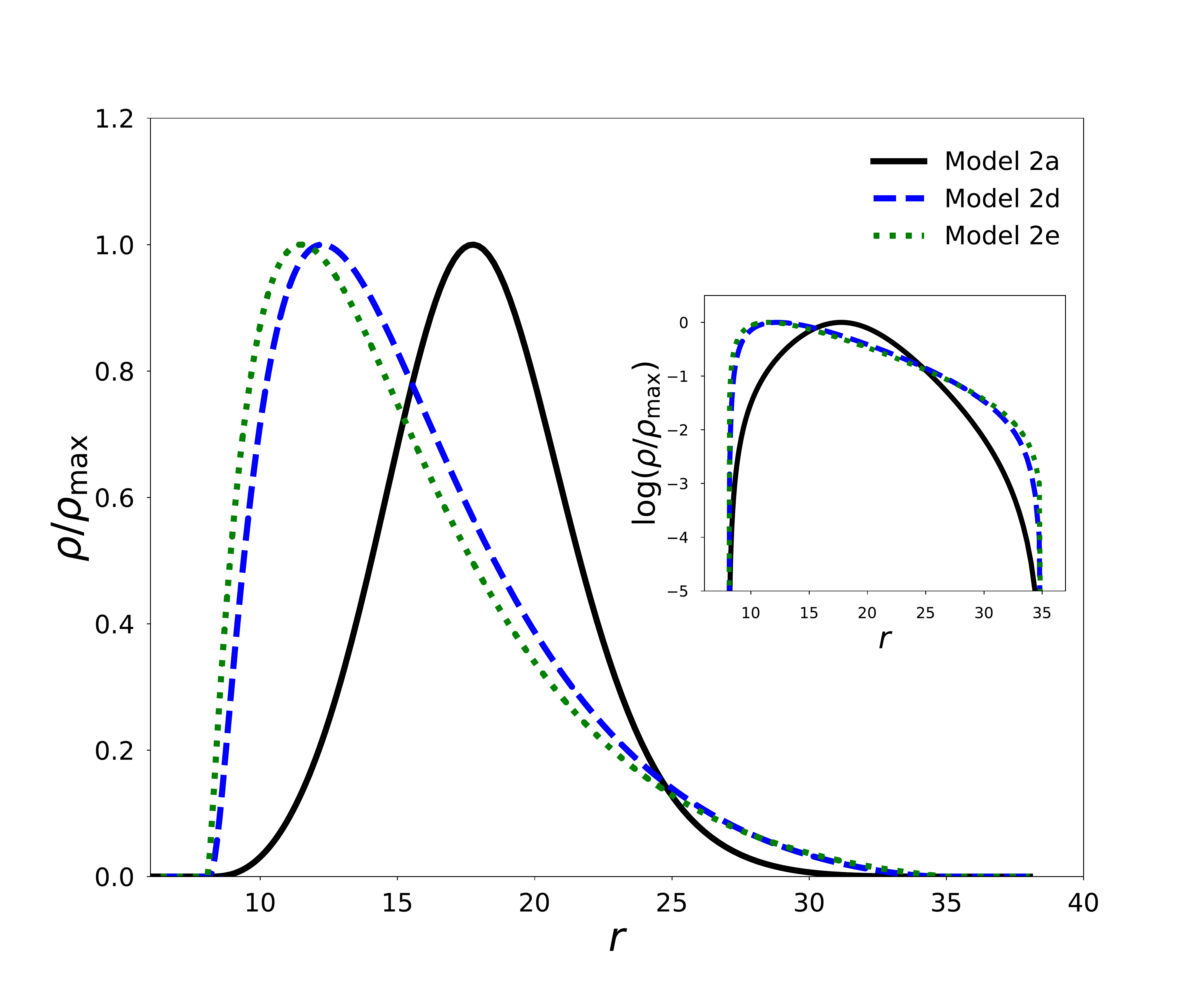} \\
\includegraphics[width=0.9\columnwidth]{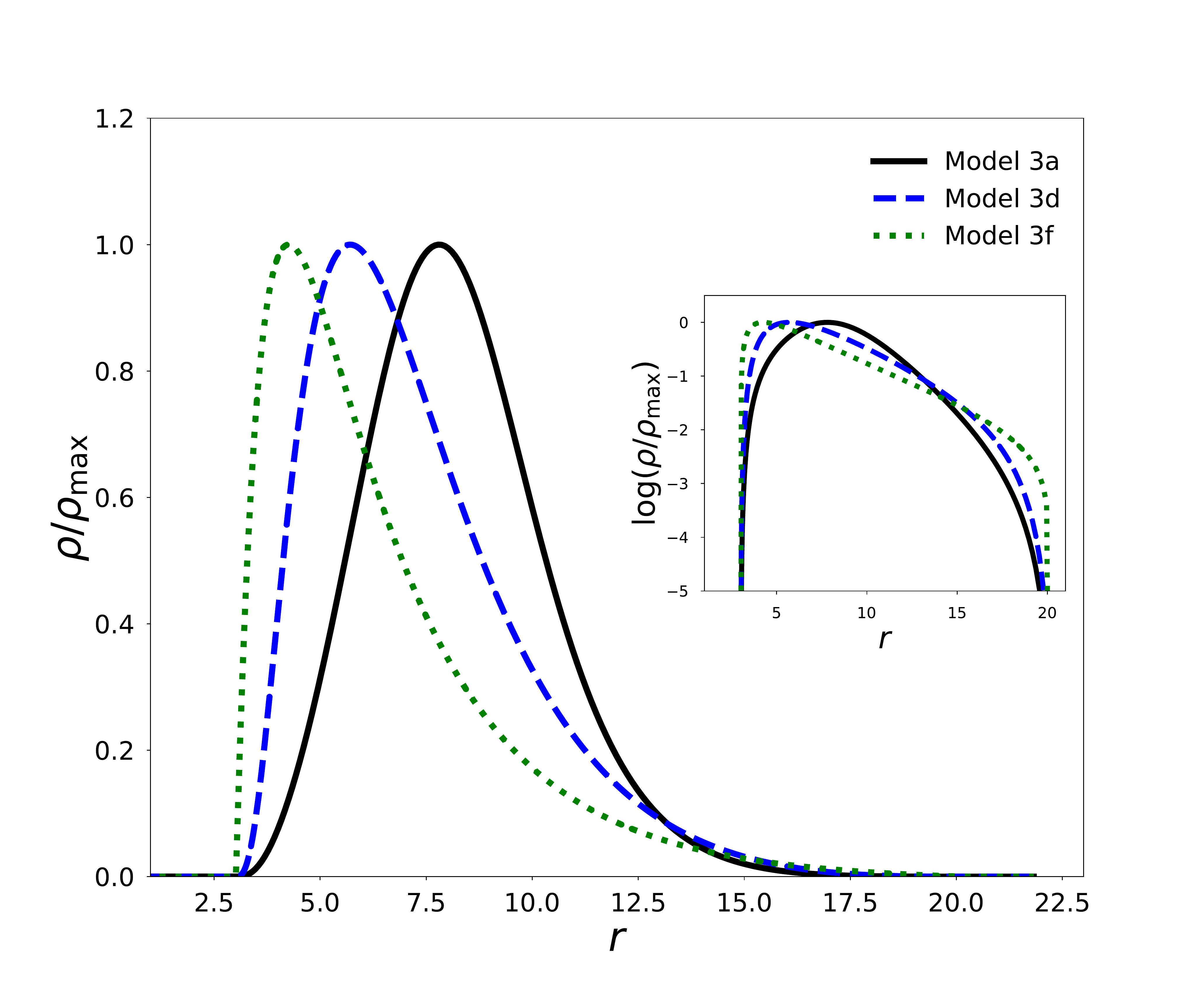}
\includegraphics[width=0.9\columnwidth]{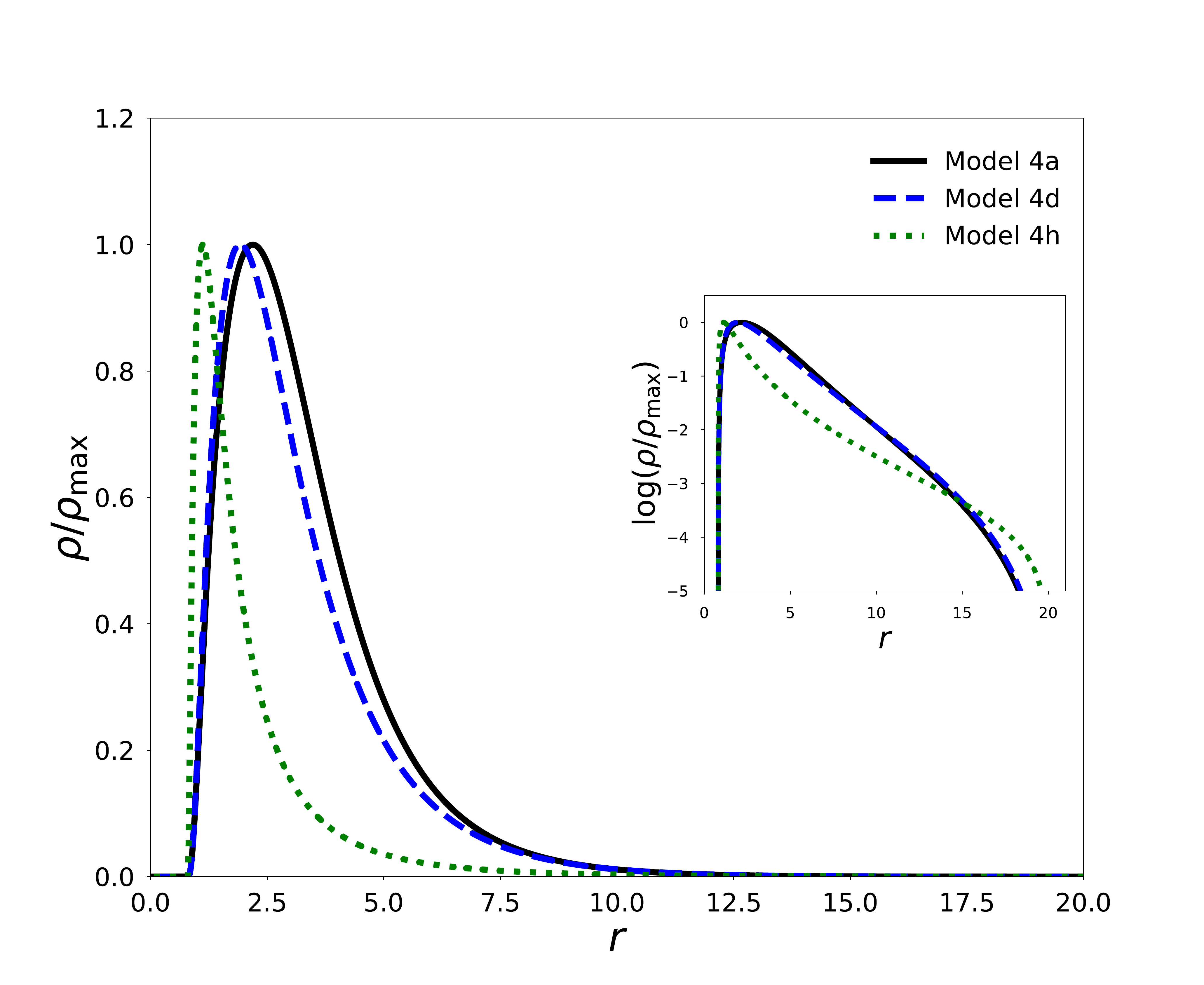}
\caption{Radial profiles of the rest-mass density at the equatorial plane for the same subset of models plotted in Fig.~\ref{models}. The insets show the same profiles in the logarithmic scale, to better account for the low-density regions.}
\label{radial_rho}
\end{figure*}

\begin{figure}
\centering
\includegraphics[width=0.9\columnwidth]{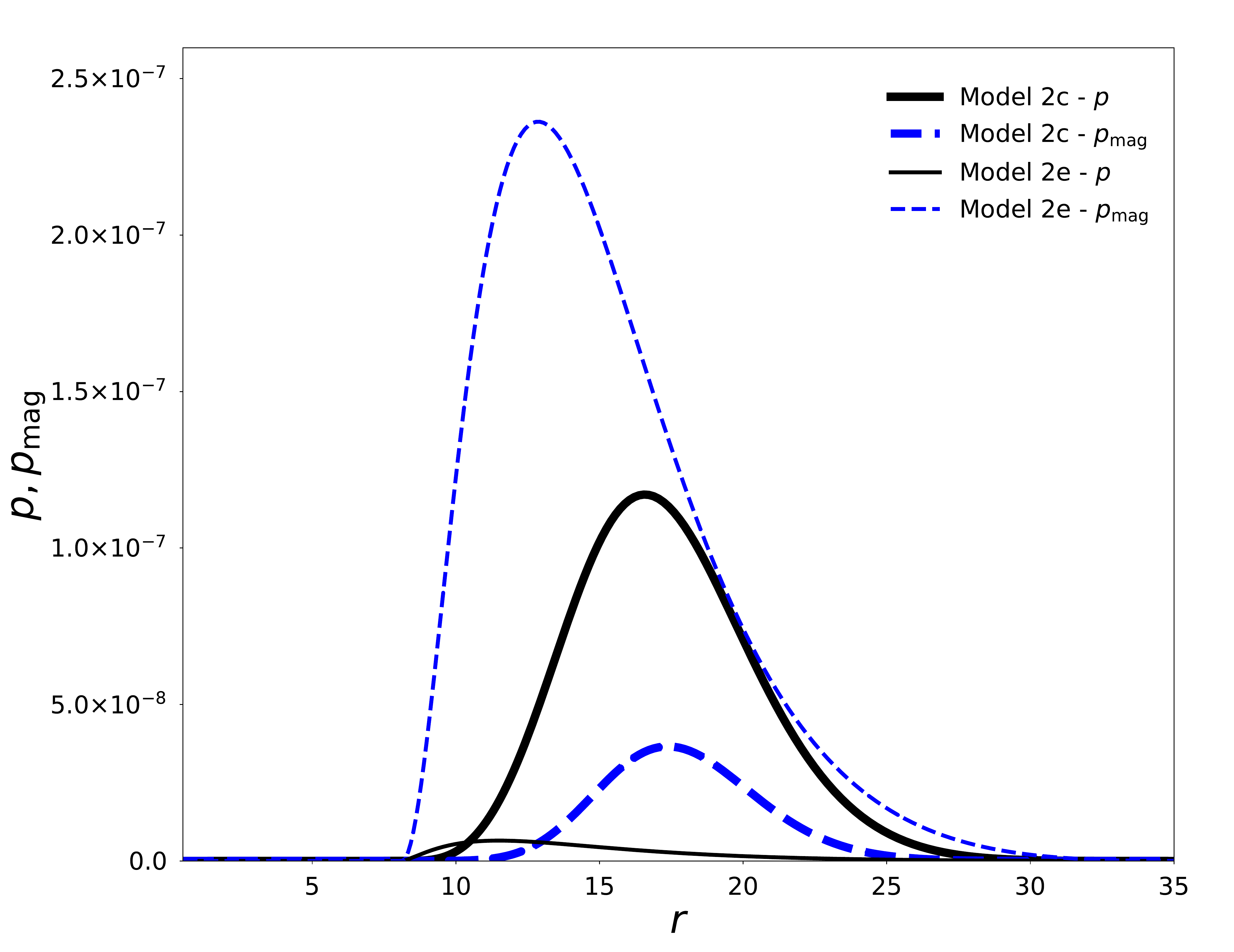} 
\\
\includegraphics[width=0.9\columnwidth]{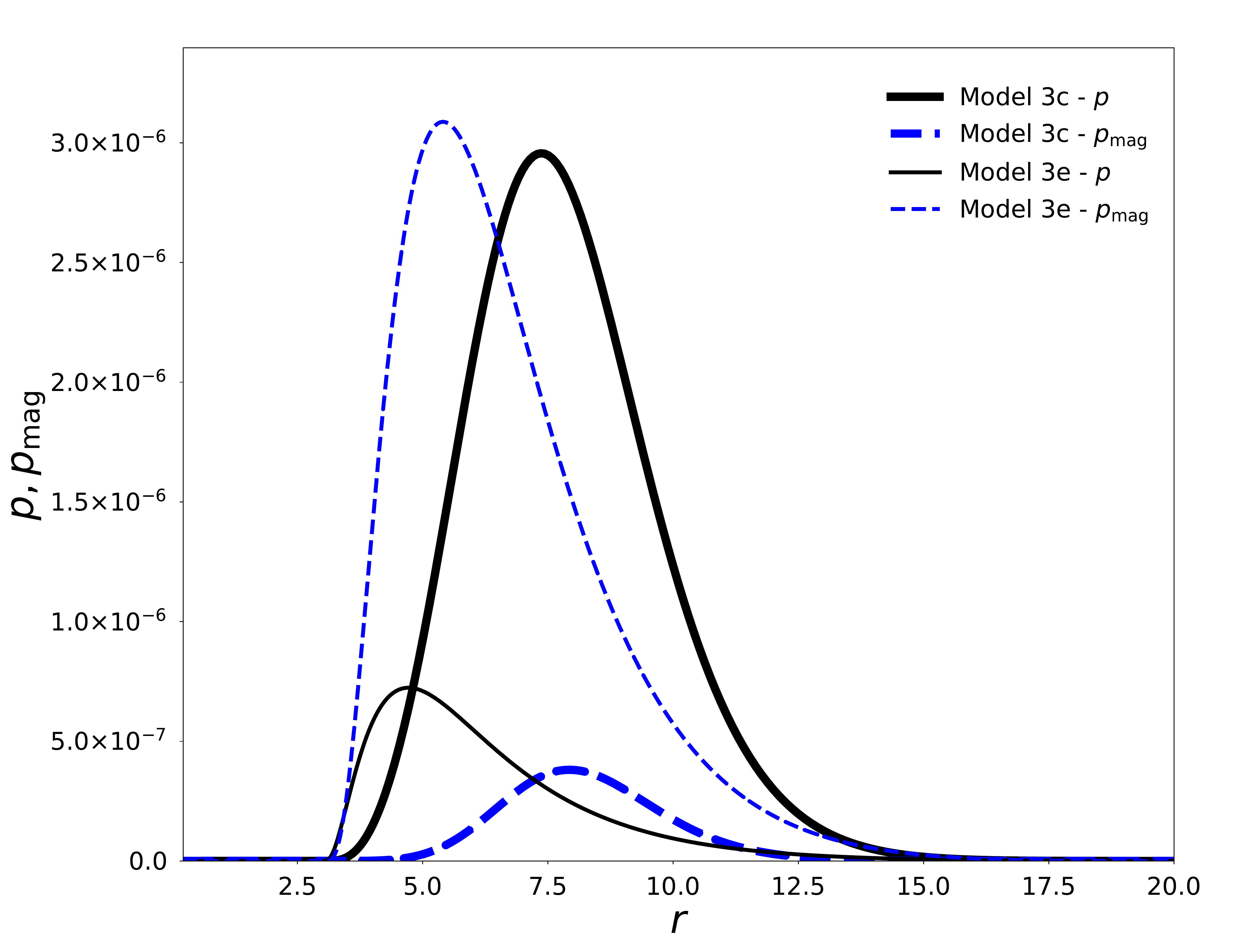}
\caption{Comparison of the thermal pressure $p$ (solid black lines) and the magnetic pressure $p_\mathrm{mag} = \frac{1}{2}b^2$ (dashed blue lines) at the equatorial plane.}
\label{magnetic_shift}
\end{figure}

\begin{figure*}
\centering
\includegraphics[width=0.9\columnwidth]{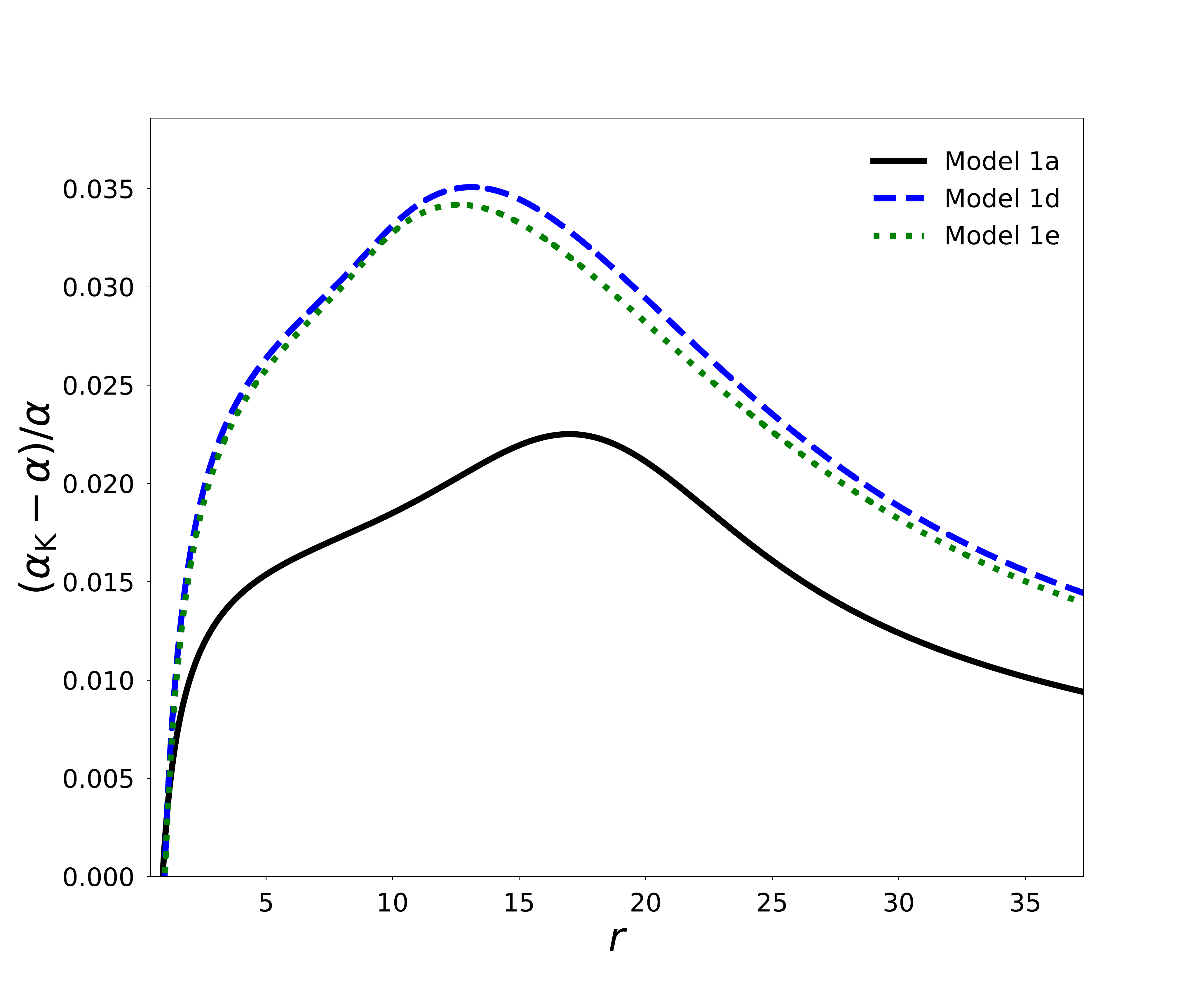}
\includegraphics[width=0.9\columnwidth]{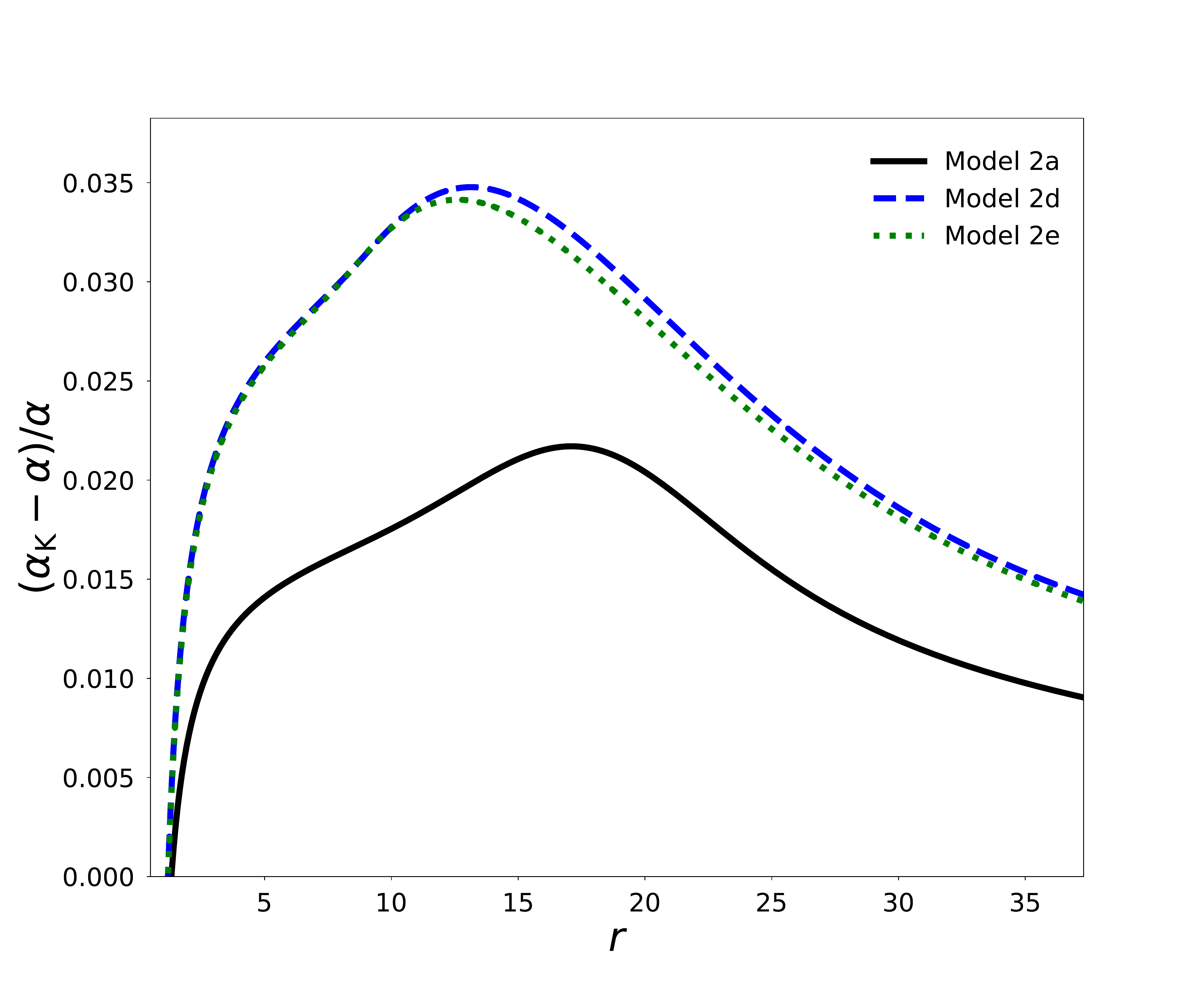} \\
\includegraphics[width=0.9\columnwidth]{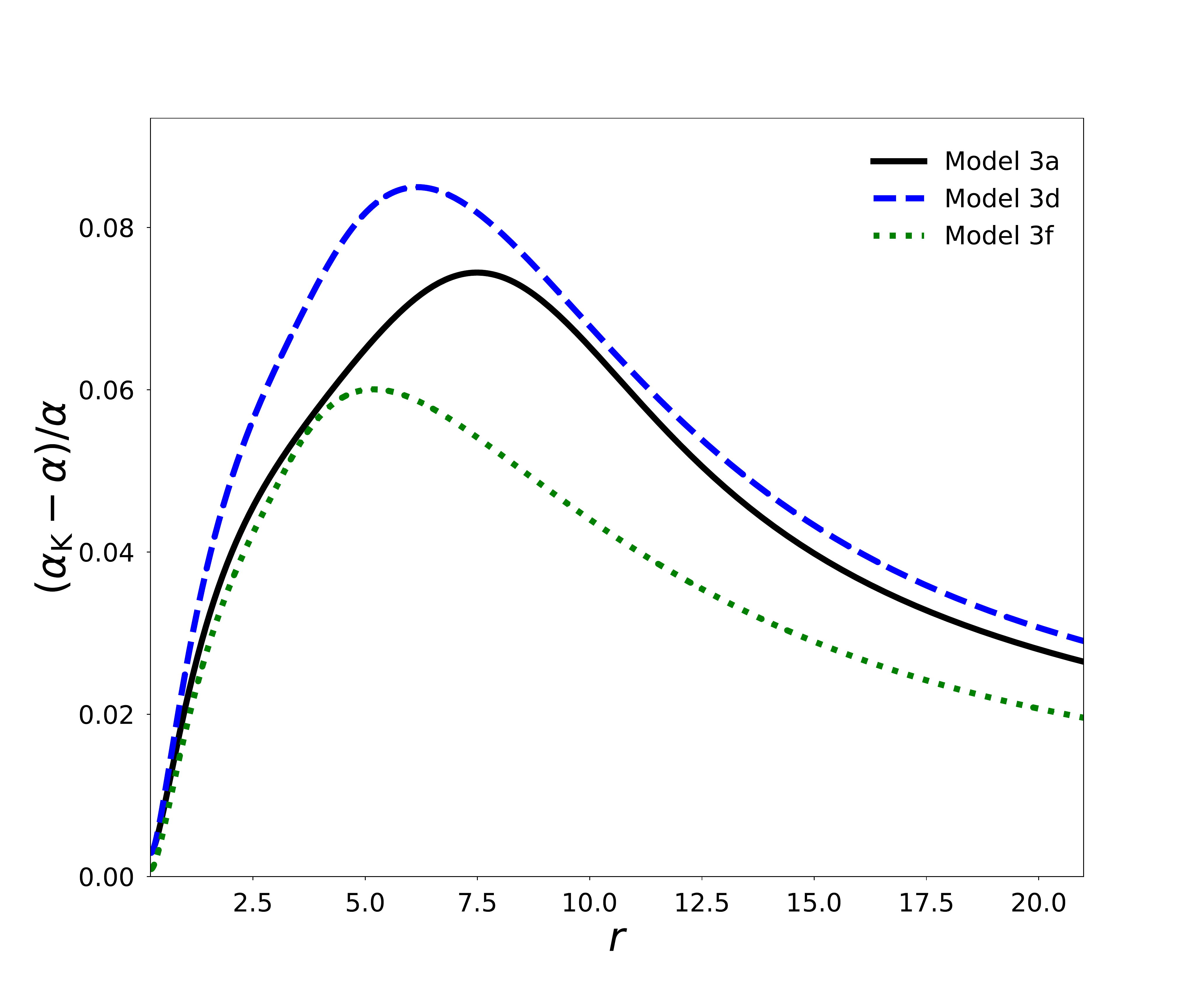}
\includegraphics[width=0.9\columnwidth]{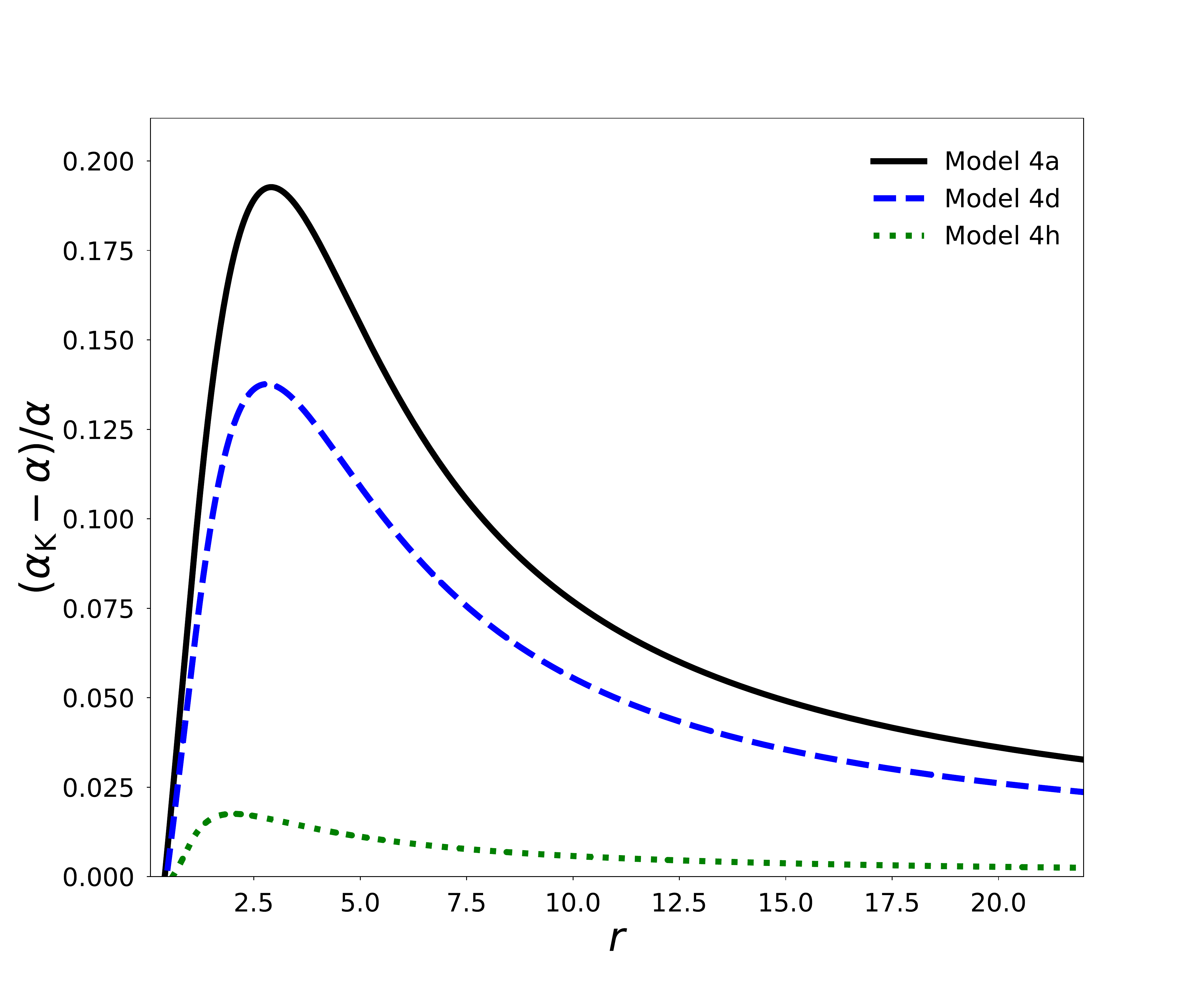}
\caption{Radial profiles of the deviation of the lapse function $\alpha$ with respect to its value for an isolated Kerr black hole with mass $m_\mathrm{ADM}$ and spin $a$. The models displayed are the same subset of models plotted in Fig.~\ref{models}.}
\label{radial_alpha}
\end{figure*}

\begin{figure*}
\centering
\includegraphics[width=0.9\columnwidth]{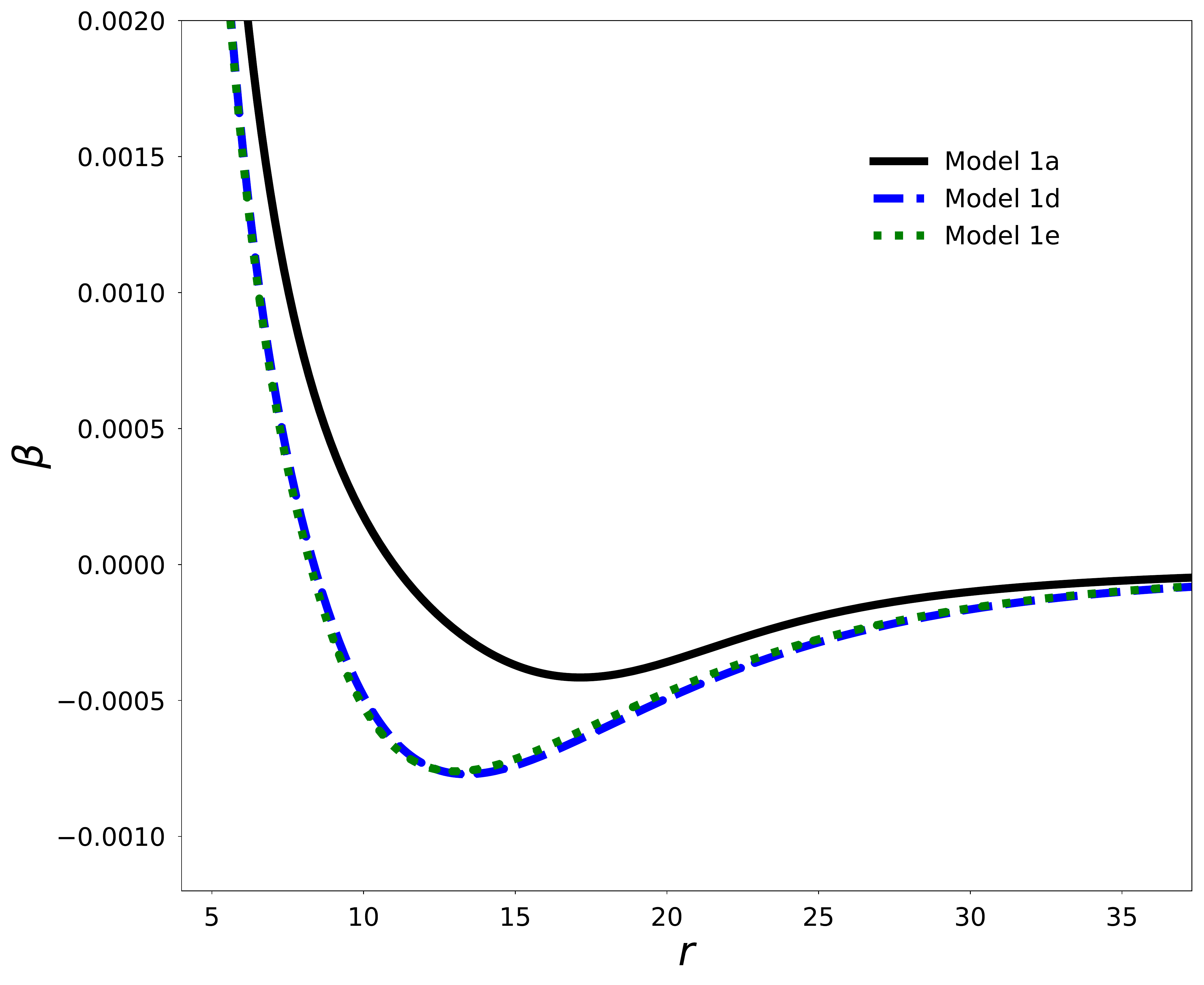}
\includegraphics[width=0.9\columnwidth]{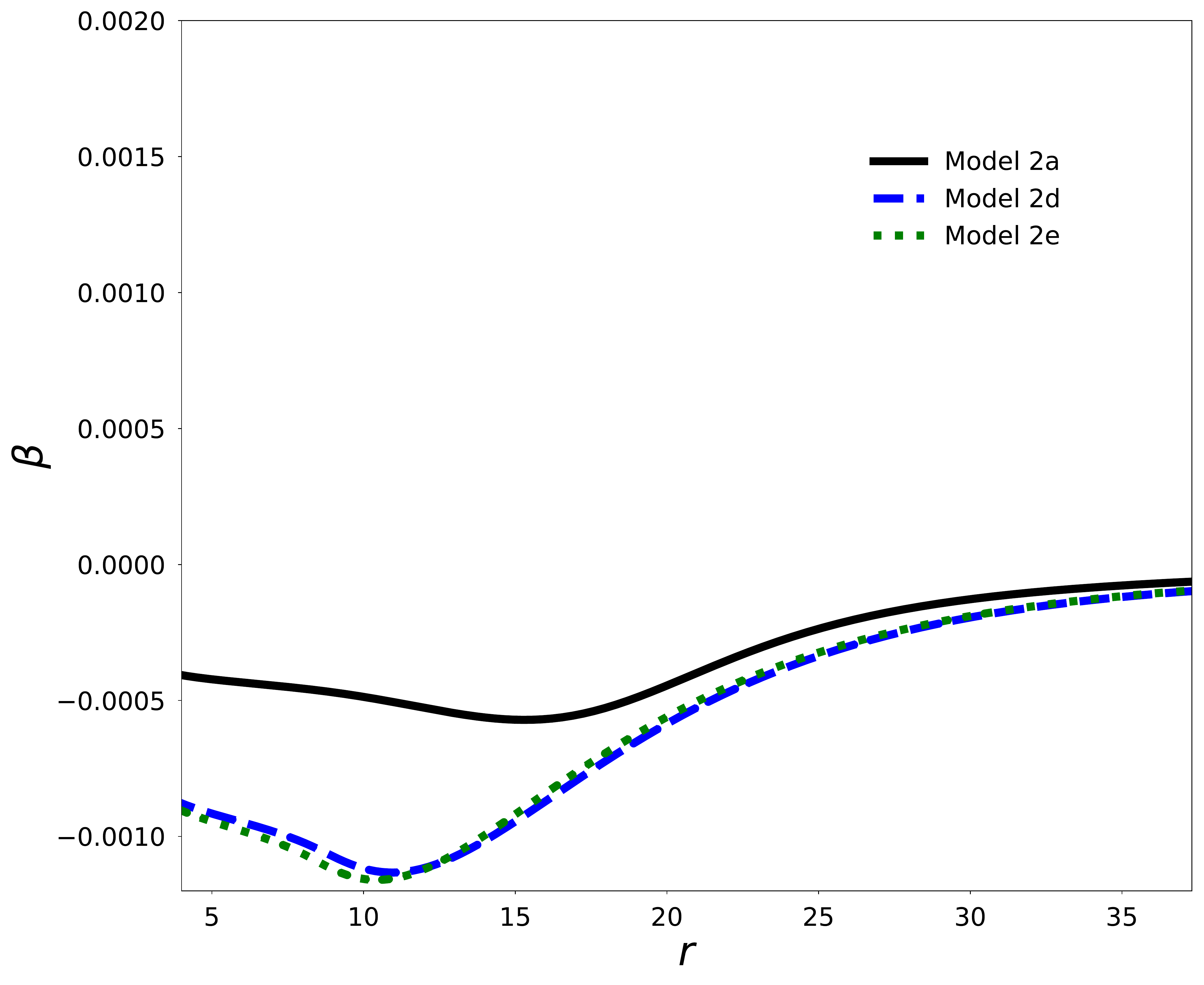} \\
\includegraphics[width=0.9\columnwidth]{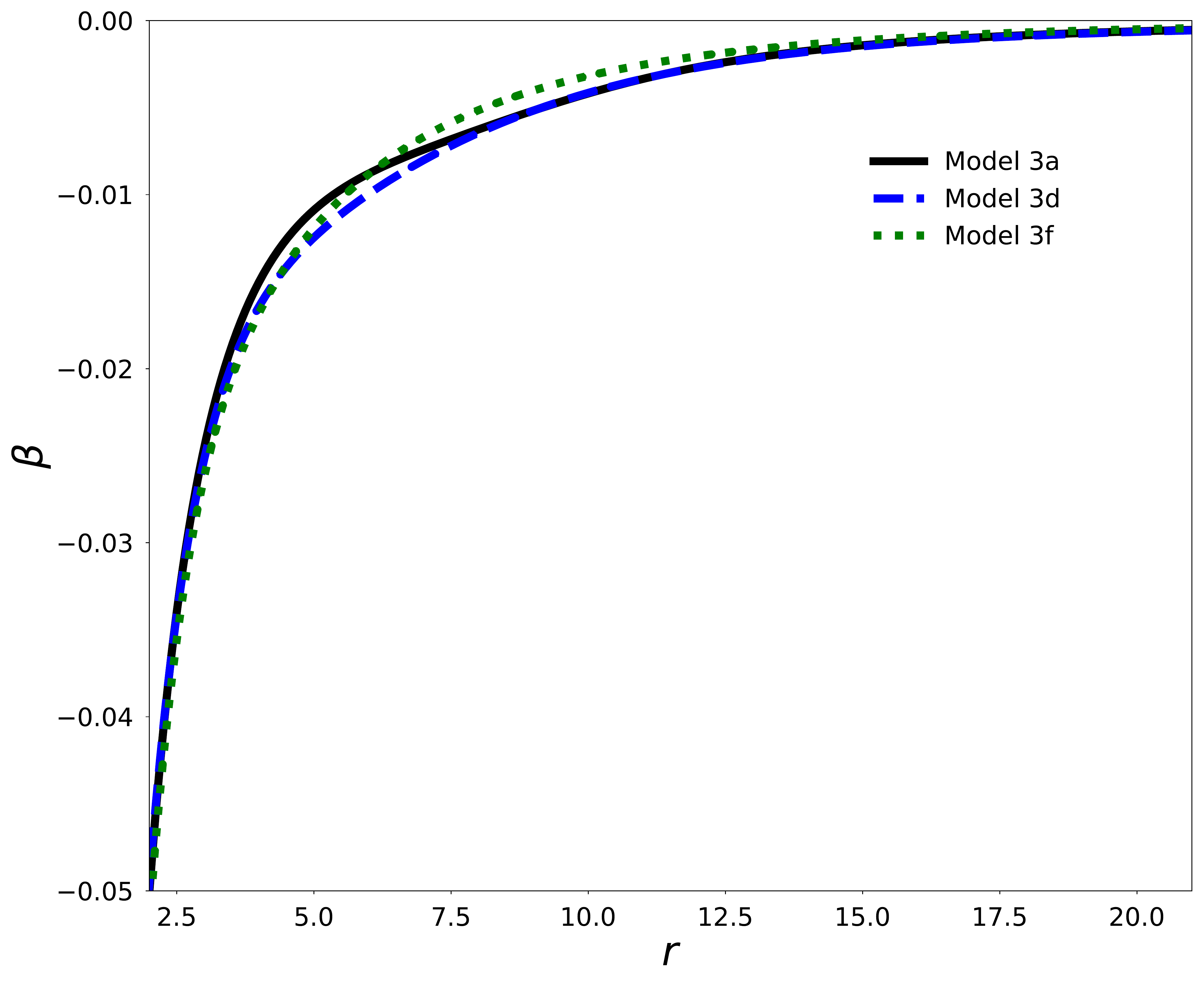}
\includegraphics[width=0.9\columnwidth]{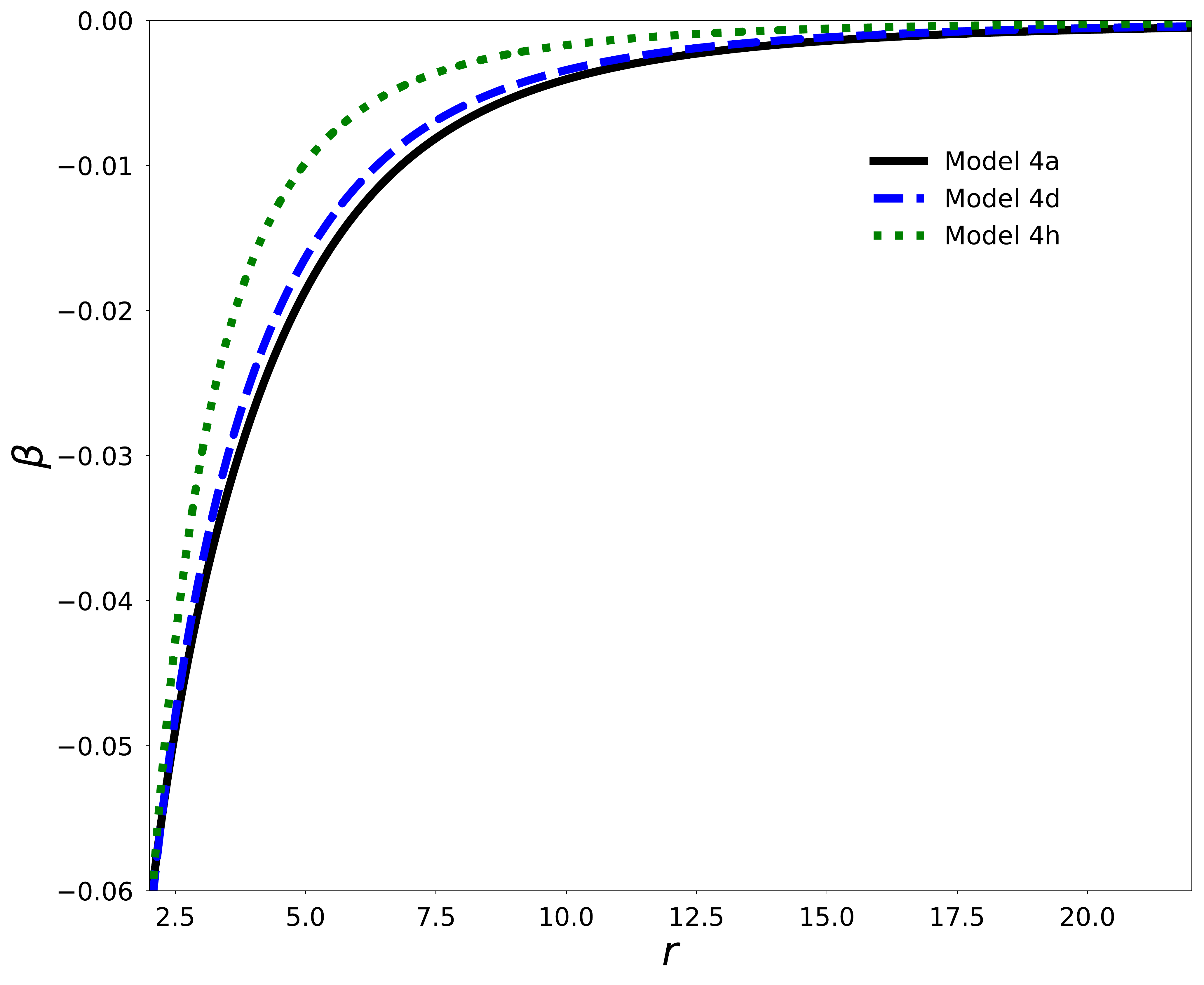}
\caption{Radial profiles of the shift vector component $\beta$ at the equatorial plane. The models displayed are the same subset of models plotted in Fig.~\ref{models}.}
\label{radial_beta}
\end{figure*}

The numerical solutions are specified by the following set of parameters: the black hole mass and angular momentum parameters, $m$ and $a$, the inner and outer radii of the disk $r_1$, $r_2$, the polytropic exponent of the equation of state $\gamma$, the maximum rest-mass density within the disk $\rho_\mathrm{max}$, and the constants $C_1$ and $n$ that characterize the prescription of the magnetic field. We note that this parameterization does not specify solutions uniquely. In fact, even in the case with no magnetic field one can observe a bifurcation: two solutions corresponding to different asymptotic masses can be obtained for fixed $m$, $a$, $r_1$, $r_2$, $\gamma$, and $\rho_\mathrm{max}$. Usually, one of these solutions corresponds to a case with the mass of the torus much larger than the mass of the central black hole \cite{kulczycki}. This effect is interesting per se, and it will be the subject of a separate study.

The values $r_1$ and $r_2$ refer to coordinate radii of the torus. The simplest way of obtaining a geometrical size measure would be to use the circumferential radius $r_\mathrm{C}$ related to $r$ at the equatorial plane by the coordinate transformation $r_\mathrm{C} = \psi^2 r$. In the following, by $r_\mathrm{C,1}$ and $r_\mathrm{C,2}$ we will denote the circumferential radii corresponding to coordinate radii $r_1$ and $r_2$, respectively. It should be kept in mind that in the strong gravitational-field regime the relation between $r$ and $r_\mathrm{C}$ may be not monotonic. In fact, numerical solutions representing self-gravitating tori with a maximum of $r_\mathrm{C}$ occurring within the torus, and not at its outer edge, were computed in \cite{labranche}.

We measure the impact of the magnetic field by computing a magnetisation parameter $\beta_\mathrm{mag}$ defined as
\begin{eqnarray} 
\beta_\mathrm{mag} = \frac{p}{p_\mathrm{mag}} = \frac{2p}{b^2}, 
\end{eqnarray}
and evaluated at the maximum of the thermal pressure $p$.

We have computed a sample of numerical solutions. Their parameters are reported in Table~\ref{table1}. This table also provides the values of a handful of quantities that can be used to characterize the solutions: the total ADM mass $m_\mathrm{ADM}$, the mass of the black hole $m_\mathrm{BH}$, the angular momentum of the torus $J_1$, the circumferential inner and outer radii of the torus $r_\mathrm{C,1}$, $r_\mathrm{C,2}$, and the magnetisation parameter $\beta_\mathrm{mag}$. In all our numerical examples we set $m = 1$. This can be viewed as fixing the system of units; in practice, setting $m = 1$ yields $m_\mathrm{BH} \approx 1$. Table~\ref{table1} is divided into parts which group models with the same values of $a$, $r_1$, $r_2$, $\rho_\mathrm{max}$ but different values of $C_1$. In particular, each group corresponds to a different value of $a$; we chose specifically $a = -0.5$, $0$, $0.9$, and $0.99$. A negative value of $a$ means counterrotation, i.e.~we adhere to a convention with $\Omega > 0$. For simplicity, we fix in all our solutions the value of the polytropic exponent to $\gamma = 4/3$ and the parameter $n$ of the magnetisation law to $n = 1$. Except for the models with the fastest spinning black hole ($a=0.99$), in all investigated cases we were able to obtain solutions with the magnetisation parameter $\beta_\mathrm{mag}$ ranging from $\infty$ (no magnetic field) to the level of the order of $10^{-3}$ to $10^{-4}$ (i.e.~highly magnetised models). The case with ($a=0.99$) listed in Table~\ref{table1} is exceptional: the tori characterized by small values of $\beta_\mathrm{mag}$ are fairly light. Moreover, a large number of numerical iterations  ($\sim 10^5$) is required in order to converge to a solution.

It is not unreasonable to assume that stable solutions should have $r_1$ larger than the location of the Innermost Stable Circular Orbit (ISCO). Because of the self-gravity of the torus, the location of the ISCO deviates from the value characteristic for the Kerr spacetime with a given mass $m$ and spin parameter $a$. Nevertheless, the Kerr values can be still used to get a rough estimate of the location of the ISCO. For $m = 1$ and $a = -0.5$, the circumferential radius of the ISCO in the Kerr spacetime is $r_\mathrm{C, ISCO} = 7.57$. For $m = 1$ and $a = 0, 0.9$ and $0.99$, we obtain, respectively, $r_\mathrm{C, ISCO} = 6, 2.63$, and $2.11$. Of course, given a numerical solution, the true location of the ISCO can also be computed, for instance using the formalism described in \cite{sasaki}. We have actually checked that the solutions listed in Table \ref{table1} satisfy the condition $r_\mathrm{C,1} > r_\mathrm{C, ISCO}$. A detailed analysis of the influence of the self-gravitating torus on the location of the ISCO will be given elsewhere.

From the inspection of all solutions listed in Table \ref{table1} we conclude that the configurations with smaller values of $\beta_\mathrm{mag}$ (relatively stronger magnetic fields) tend to have the maxima of the density shifted towards smaller radii. This behavior is illustrated in Fig.\ \ref{models}, which depicts the morphology of a subset of models by plotting the logarithm of their rest-mass density $\rho$ in the meridional half-plane. Figure \ref{radial_rho}, in which we plot radial profiles of the density at the equatorial plane, shows this trend in a more clear way. For clarity, Fig.\ \ref{radial_rho} displays the profiles both in linear and logarithmic scales (the latter in the insets). We note that the shift of the maximum of the density towards smaller radii in magnetised disks has already been observed in the test-fluid models built in \cite{gf}.

On the other hand, the analysis of the solutions with different values of the constant $C_1$ appearing in Eq.\ (\ref{bernoulli2}) shows that the larger $C_1$ the smaller the thermal pressure $p$, even in cases in which the maximum of the baryonic density $\rho$ is fixed. In general we also observe an increase of the absolute values of the magnetic pressure $p_\mathrm{mag} = \frac{1}{2}b^2$. Both factors lead to a rapid decrease of the magnetisation parameter $\beta_\mathrm{mag}$. In some sense, with an increase of $C_1$, the role of the (gradient of the) thermal pressure in counter-balancing gravity is taken over by the magnetic pressure. This effect is illustrated in Fig.\ \ref{magnetic_shift} for models 2c, 2e, 3c and 3e. Note that in both cases the maximum of the largest of the two pressures (thermal or magnetic) is about two orders of magnitude smaller than the maximum of the rest-mass density.

The presence of a magnetic field affects the total ADM mass of the system (mainly by influencing the mass of the torus) in a nontrivial way. Although a direct contribution of the terms related with the magnetic field to the mass, as computed e.g.\ from Eq.\ (\ref{torus_mass}), is small, the magnetic field changes the total mass of the system by affecting the distribution of the rest-mass density within the disk. The nature of the changes of the ADM mass with an increasing magnetic field contribution depends on the spin of the black hole and on the distance between the black hole and the torus (mainly on the location of the inner edge of the torus $r_1$). Disentangling these two factors is difficult, since the location of the ISCO depends predominantly on the spin of the black hole, and we want our models to satisfy the condition $r_\mathrm{C,ISCO} < r_\mathrm{C,1}$. The dependence of the ADM mass with the black hole spin in the examples collected in Table \ref{table1} is as follows: for $a = - 0.5$ the ADM mass grows with increasing $C_1$, for $a = 0$ and $a = 0.9$ the behaviour of the ADM mass is not monotonic with $C_1$, and for  $a = 0.99$ the ADM mass decreases with increasing $C_1$. In the latter case this effect is strong. The ADM mass drops from $m_\mathrm{ADM} = 1.70$ for $C_1 = 0$ (no magnetic field) to $m_\mathrm{ADM} = 1.05$ for $C_1 = 4.7$ (magnetisation parameter $\beta_\mathrm{mag} = 2.28 \times 10^{-2}$).

The importance of the effects of the disk self-gravity can be estimated by plotting the deviation of the lapse function at the equatorial plane with respect to its value for an isolated Kerr black hole, $(\alpha_{\mathrm{K}} - \alpha)/ \alpha$ (here $\alpha_{\mathrm{K}}$ is computed using Eq.~\eqref{eq:alpha_K}). These radial profiles are plotted in Fig.~\ref{radial_alpha} for the same subset of models depicted in Fig.~\ref{models}. We observe a correlation of the maximum deviations with the rest-mass density $\rho_{\mathrm{max}}$ of the models, the deviations being larger the larger the value of $\rho_{\mathrm{max}}$, namely $~20\%$ for model 4a, $~8.5\%$ for model 3d and $~3.5\%$ for models 1d and 2d. Furthermore, it can be seen that, as expected, the deviation grows if the fraction of the mass stored at the torus is greater. This fraction can be easily inferred from Table \ref{table1}. 

A close inspection of plots in Figs.\ \ref{radial_rho} and \ref{radial_alpha} reveals a somewhat unexpected similarity of certain features of models 1a--1f and 2a--2f. These two families of models are characterized by the same values of coordinate radii $r_1$ and $r_2$, and the same maximal densities $\rho_\mathrm{max}$, but they differ in the assumed values of the black hole spin parameter ($a = -0.5$ and $a = 0$, respectively). The plots of the rest-mass density shown in Fig.\ \ref{radial_rho} and corresponding to models belonging to both families are nearly indistinguishable from one another. One can also hardly spot any difference between models 1a--1f and 2a--2f in the plots of the differences between the lapse functions (the actual lapse and the lapse corresponding to the Kerr metric) shown in Fig.\ \ref{radial_alpha}. However, both classes of models are actually different. The differences stand out clearly in the plots of the shift vector $\beta$ shown in Fig.\ \ref{radial_beta}. Of course, the absolute values of the lapse function characterizing the models belonging to both classes are also different.

\section{Summary}
\label{summary}

We have presented general-relativistic models of stationary, axisymmetric, self-gravitating, magnetised disks (or tori) rotating around spinning black holes. They have been obtained by solving numerically the coupled system of the Einstein equations and the equations of ideal GRMHD. The mathematical formulation of our approach has closely followed the work of Shibata~\cite{shibata}, who built purely hydrodynamical self-gravitating, constant angular momentum tori around black holes in the puncture framework. The inclusion of magnetic fields represents the first new ingredient of our approach. On the other hand, building on previous studies of configurations with no magnetic field~\cite{kkmmop,kkmmop2,kmm}, we have constructed our magnetised models assuming a Keplerian rotation law in the disks, which departs from the constant angular momentum disks reported by~\cite{shibata}. The use of the Keplerian rotation law is the second new ingredient of our procedure. We have focused on toroidal distributions of the magnetic field and presented a large set of models corresponding to a wide range of values of the magnetisation parameter, starting with weakly magnetised disks and ending at configurations in which the magnetic pressure dominates over the thermal one.

The impact of the magnetic field on the disk structure is mainly related to the magnetic pressure. In all investigated models, we have observed a shift of the location of the maximum of the rest-mass density towards the central black hole. The impact of the magnetisation on the total mass of the system (or the mass of the disk) depends on the black hole spin and on the geometry of the disk. It is possible to obtain classes of solutions in which the mass of the torus decreases with the decreasing magnetisation parameter, but the converse can also be true.

All our solutions have been obtained for the polytropic equation of state with the polytropic exponent $\gamma = 4/3$, and for a specific choice of the magnetisation law. These choices can, of course, have an impact on the obtained results. Furthermore, the values of the ratio of the black-hole mass to the total mass of the system reported in this paper were kept within a reasonable range: the obtained disks are massive enough so that the effects connected with the self-gravity become important. On the other hand, in this work we have not considered disks with masses exceeding the mass of the central black hole. It is known that allowing for sufficiently large disk masses can lead to the occurrence of several general-relativistic effects, characteristic of the strong gravitational-field regime. In \cite{ansorg:2006} Ansorg and Petroff showed that a perfect fluid torus rotating (rigidly) around a black hole can create its own ergoregion, disconnected from the ergoregion of the black hole. (This effect is also present in more exotic objects, for instance the scalar hairy black holes described in~\cite{Herdeiro15}, for which the scalar field has a toroidal distribution.) In \cite{labranche} Labranche, Petroff, and Ansorg gave examples of perfect fluid tori (with no central object) in which the circumferential radius attains its maximum inside the torus, and not at its outer edge. We expect these effects to be present also within our formulation.

The results presented in this paper can be extended in several directions. On the one hand we plan to investigate the influence of the self-gravity of the torus on the location of the ISCO of a rotating black hole. In addition, we will also  study the non-linear stability properties of our solutions through numerical-relativity simulations in a dynamical spacetime setup. There are a number of instabilities that may affect the disks, such as the runaway, the Papaloizou-Pringle and the magneto-rotational instabilities. In particular, the development of the PPI in tori threaded by toroidal magnetic fields may be significantly affected by the concurrent development of the MRI, as shown recently by~\cite{bugli} for non-self-gravitating disks. The self-gravitating, magnetised tori we have built in this work can be used  to investigate the generality of those findings beyond the test-fluid limit.

The presented solutions, and models of self-gravitating magnetised disks in general, should be also relevant to the ongoing attempts to estimate the amount of angular momentum within a given volume. This is an interesting area of research within  mathematical relativity \cite{dain, khuri}. Recent works focusing on such estimates for stationary Keplerian self-gravitating disks around black holes include \cite{kmm, kmmpx}.


\begin{acknowledgments}
PM was partially supported by the Polish National Science Centre grant No.\ 2017/26/A/ST2/00530. JAF acknowledges support from the Spanish MINECO (grant AYA2015-66899-C2-1-P) and from  the  European  Union's  Horizon  2020  RISE programme H2020-MSCA-RISE-2017 Grant No.~FunFiCO-777740. MP was supported in part by ``PHAROS'', COST Action CA16214; LOEWE-Program in HIC for FAIR.

\end{acknowledgments}

\appendix*
\section{Kerr metric in quasi-isotropic coordinates}

For completeness, we express the Kerr metric in the quasi-isotropic coordinates of the form (\ref{isotropic}) \cite{shibata,brandtseidel}.

Define
\begin{eqnarray}
r_\mathrm{K} & = & r \left( 1 + \frac{m}{r} + \frac{m^2 - a^2}{4 r^2} \right), \\ 
\Delta_\mathrm{K} & = & r_\mathrm{K}^2 -2r_\mathrm{K}+a^2, \\
\Sigma_\mathrm{K} & = & r_\mathrm{K}^2 + a^2 \cos^2 \theta.
\end{eqnarray}

The Kerr metric can be expressed as
\begin{eqnarray}
g & = & - \alpha_\mathrm{K}^2 dt^2 + \psi^4 e^{2q_\mathrm{K}} (dr_\mathrm{K}^2 + r_\mathrm{K}^2 d\theta^2) + \nonumber \\
&& \psi_\mathrm{K}^4 r_\mathrm{K}^2 \sin^2 \theta (\beta_\mathrm{K} dt + d \varphi)^2,
\end{eqnarray}
where
\begin{eqnarray}
\psi_\mathrm{K} & = & \frac{1}{\sqrt{r}}\Bigl( r^2_\mathrm{K}  +a^2 +2ma^2\frac{r_\mathrm{K}\sin^2\theta  }{\Sigma_\mathrm{K}}\Bigr)^{1/4}, \\
\beta_\mathrm{K} & = & -\frac{2mar_\mathrm{K}}{(r^2_\mathrm{K}+a^2)\Sigma_\mathrm{K} +2ma^2r_\mathrm{K} \sin^2\theta}, \\
\alpha_\mathrm{K} & = & \left[ \frac{ \Sigma_\mathrm{K} \Delta_\mathrm{K}}{(r_\mathrm{K}^2+a^2)\Sigma_\mathrm{K}+2ma^2r_\mathrm{K}}\sin^2\theta \right]^{1/2}, \label{eq:alpha_K}\\
e^{q_\mathrm{K}} & = & \frac{\Sigma_\mathrm{K}}{\sqrt{(r^2_\mathrm{K}+a^2)\Sigma_\mathrm{K} +2ma^2r_\mathrm{K} \sin^2\theta}}.
\end{eqnarray}
The functions $H_\mathrm{E}$ and $H_\mathrm{F}$ corresponding to the Kerr metric read
\begin{eqnarray}
H_\mathrm{E} & = & \frac{ma \left[ (r_\mathrm{K}^2 - a^2) \Sigma_\mathrm{K} + 2 r_\mathrm{K}^2 (r_\mathrm{K}^2 + a^2) \right]}{\Sigma_\mathrm{K}^2}, \\
H_\mathrm{F} & = & - \frac{2 m a^3 r_\mathrm{K} \sqrt{\Delta_\mathrm{K}} \cos \theta \sin^2 \theta}{\Sigma_\mathrm{K}^2}.
\end{eqnarray}

\end{document}